\documentclass[journal,a4paper,10pt]{IEEEtran}
\usepackage{amsmath,amssymb,amsfonts} 
\usepackage{graphicx}                 
\usepackage{fontenc}
\usepackage{multirow} 
\usepackage{theorem}
\usepackage{array}

\usepackage[footnotesize]{caption}
\usepackage{color,cite}
\usepackage{algorithmic}
\usepackage{algorithm}
\usepackage{epstopdf}
\usepackage{mathrsfs}
\usepackage[english]{babel}
\usepackage{fancyhdr}
\usepackage[inline]{enumitem}

\usepackage{graphicx}
\graphicspath{{figs/}}
\DeclareGraphicsExtensions{.pdf,.jpeg,.png,.jpg,.eps,.PNG,.bmp}
\usepackage{subfig}
 
\makeatother

\ifodd 1
\newcommand{\rev}[1]{{\color{black} #1}} 
\newcommand{\revtwo}[1]{{\color{black} #1}} 
\else

\fi

\begin{document}
\title{\rev{Understanding Urban Human Mobility through Crowdsensed Data}}

\author{Yuren Zhou,~Billy Pik Lik Lau,~Chau Yuen,~Bige Tun\c{c}er,~Erik Wilhelm

\thanks{Y. Zhou, B. P. L. Lau, C. Yuen, B. Tun\c{c}er, and E. Wilhelm are with the Singapore University of Technology and Design (SUTD), 8 Somapah Road, Singapore 487372 (Email: \{yuren\_zhou, billy\_lau\}@mymail.sutd.edu.sg, \{yuenchau, bige\_tuncer, erik\_wilhelm\}@sutd.edu.sg).}
}
\IEEEoverridecommandlockouts
\maketitle
\thispagestyle{fancy}
\begin{abstract}

Understanding how people move in the urban area is important for solving urbanization issues, such as traffic management, urban planning, epidemic control, and communication network improvement. Leveraging recent availability of large amounts of diverse crowdsensed data, many studies have made contributions to this field in various aspects. They need proper review and summary. In this paper, therefore, we first review these recent studies with a proper taxonomy with corresponding examples. Then, based on the experience learnt from the studies, we provide a comprehensive tutorial for future research, which introduces and discusses popular crowdsensed data types, different human mobility subjects, and common data preprocessing and analysis methods. Special emphasis is made on the matching between data types and mobility subjects.
\rev{Finally, we present two research projects as case studies to demonstrate the entire process of understanding urban human mobility through crowdsensed data in city-wide scale and building-wide scale respectively}.
Beyond demonstration purpose, the two case studies also make contributions to their category of certain crowdsensed data type and mobility subject.

\end{abstract}
\begin{IEEEkeywords}
Human mobility, Mobile crowdsensing, Big data analysis, Urban computing, WiFi-based tracking.
\end{IEEEkeywords}

\section{Introduction}

Understanding urban human mobility is understanding how people move in the urban area in various aspects, such as commuting distance, connectivity between people and places, and commonly used transportation. It is therefore very helpful for solving urban issues. For example, estimating human flows between different origins and destinations helps arrange better public transportation~\cite{demissie2016inferring}; studying contacts between residents on their daily routes helps simulate the dynamics of disease transmission~\cite{vazquez2013using};
and understanding patterns of human movements and connectivity helps establish opportunistic networks which improves the connectivity of mobile devices~\cite{karamshuk2011human}.


Early studies of human mobility mainly relied on census data, so it was difficult to obtain comprehensive results due to limitations of data amount and coverage. Nowadays, with the help of well-developed mobile communication technology, it is possible to collect large-volume, diverse, and fine-grained data related to human mobility. The research of human mobility thus becomes heavily data-driven and spreads into various subjects (directions). For this nature, recent studies can be categorized based on the types of data they analyze and the mobility subjects they study.

Despite many sources of human mobility data, we focus on crowdsensed data. Crowdsensing means collecting data from ubiquitous mobile devices carried by people, so it generally has larger volume, more geographic coverage, and better dynamics, compared with census and fixed traffic sensors. According to different data collection mechanisms, common crowdsensed mobility data can be summarized into seven types,
as shown in the two side columns of Table~\ref{table:literature_review}.

Based on the scale, urban human mobility subjects can be separated into two classes, city-wide and building-wide. The former describes the way people travel in the city-level scope, while the latter is focused on how people move between near-by buildings or in a large structure.
\rev{As the field is very active, here we limit our review coverage to four most common city-wide mobility subjects and two building-wide subjects which investigate urban human mobility through crowdsensed data, as shown in the second row of Table~\ref{table:literature_review}}.

\begin{table*}[!h]
\centering
\caption{\textbf{Common crowdsensed mobility data types and mobility research subjects with their possible combinations}, supported by representative examples. M.1-7 stand for data analysis methods introduced in section \textit{Research Steps and Analysis Methods}. The missing content of a block means improper combination of the corresponding data type and mobility subject.}
\includegraphics[width=\linewidth]{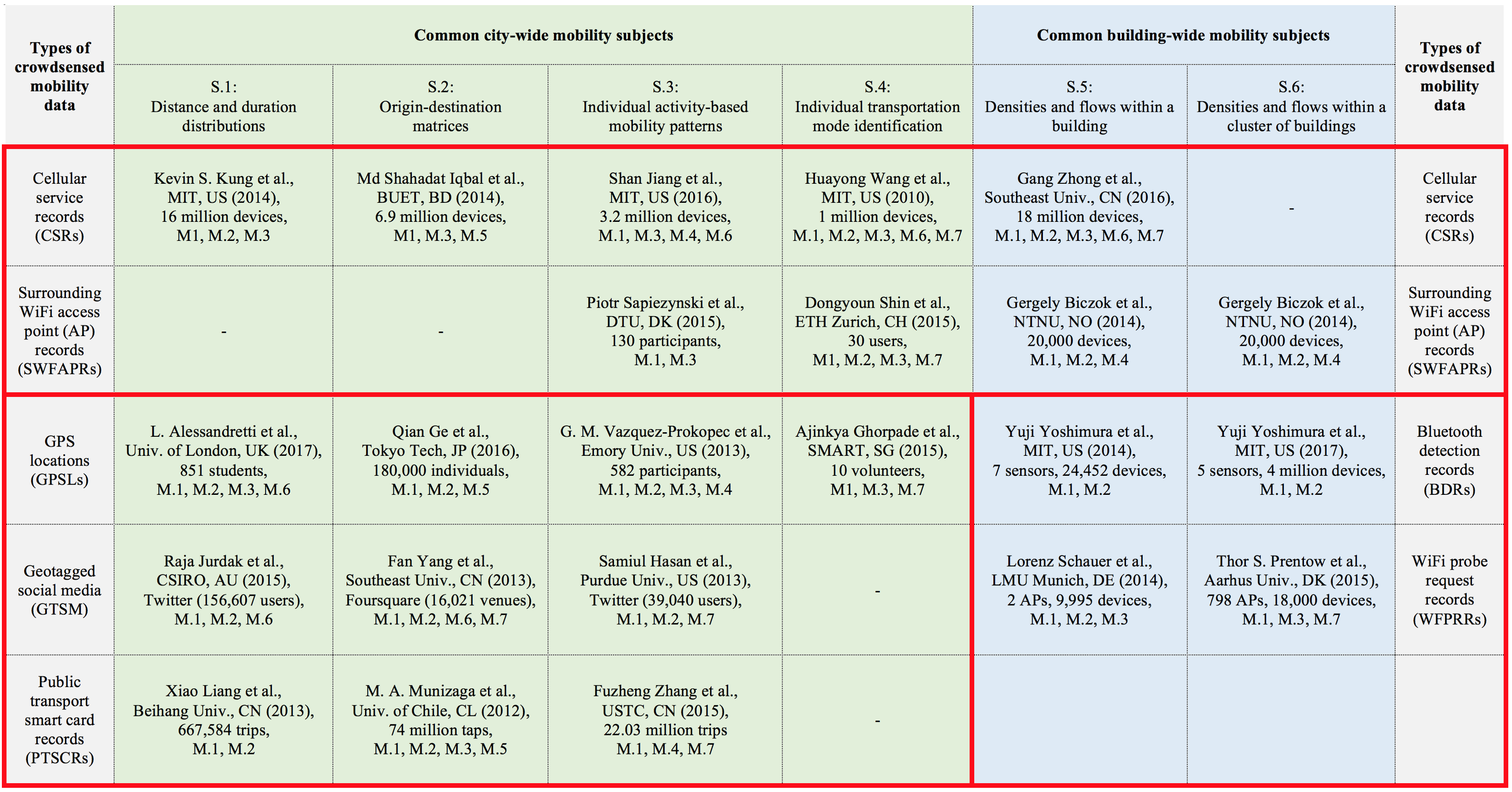}
\label{table:literature_review}

\end{table*}

In practice, not every pair of data type and mobility subject is compatible. In Table~\ref{table:literature_review}, we show possible combinations of the data types and mobility subjects supported by representative example studies. 
From the table, one can also find that different data types have different strengths in terms of scale and precision, even if they are suitable for the same subject.

As a result, it is critical to study suitable mobility subject when given certain data type, or to select the best data type when having certain mobility question. However, such selection problems have not been well discussed. This motivates the tutorial part of this paper which guides readers to make proper selection by discussing:
\begin {enumerate*}
\item pros and cons of each crowdsensed mobility data type,
\item details of each human mobility subject, 
\item suitability of each data type for each mobility subject,
\item common data preprocessing and analysis methods. 
\end {enumerate*}

To further demonstrate the process of understanding urban human mobility through crowdsensed data, two different-scale research projects are presented as case studies.
\revtwo{In the city-wide case study, we collected surrounding WiFi access point (AP) records (SWFAPRs) to discover individual daily activity and transportation usage patterns. The existing studies have  two common weaknesses. Firstly, the data collection scale is too small to achieve city-wide findings, and secondly, the results are not obtained purely from SWFAPRs but demanding other data types.
Our work, in contrast, deployed a much larger-scale data collection (around 90,000 participants) and obtains precise patterns from only SWFAPRs.
In the building-wide case study, WiFi probe request records (WFPRRs) were collected to analyze pedestrian visiting and moving behaviors between nearby buildings. 
We are able to obtain precise information about pedestrians (e.g. walking path and stay time in buildings) in a passive way making no influence on people. Compared with others, this work studies a complex and busy district consisting of shopping malls, office buildings, and a transportation hub.
}

\section{Data Types and Mobility Subjects}

\subsection{Types of Crowdsensed Data}

\rev{In this subsection, we explain the seven crowdsensed mobility data types, discuss their pros and cons, and summarize their applicable mobility subjects. S.1-6 stand for the six mobility subjects as shown in Table~\ref{table:literature_review}.}

\textbf{Cellular Service Records (CSRs):} CSRs are records of mobile phones' activities collected by telecommunication companies. Because they contain timestamps of the activities and identifier of the serving cell stations, CSRs can provide spatiotemporal information of mobile phone users. Due to wide ownership of mobile phones,
\rev{CSRs have superiorities in geographic and demographic coverage, data volume (millions in magnitude as shown in Table~\ref{table:literature_review})},
and collection cost.
\rev{In city area, space between cell towers varies from 50 meters to several kilometers~\cite{jiang2017activity}.}
\rev{From city-wide point of view, CSRs still have drawbacks of large spatial granularity (from hundreds to thousands of meters),}
poor temporal continuity, and frequent location jumping (due to handover between cell towers).
\rev{According to Table~\ref{table:literature_review}, CSRs can be applied to five mobility subjects: S.1-5.}

\textbf{Surrounding WiFi AP Records (SWFAPRs):} SWFAPRs can be collected by common mobile devices by scanning surrounding WiFi APs. Due to dense WiFi coverage and unique MAC address of WiFi AP, localization service based on SWFAPRs is commercially available (e.g. Skyhook Wireless). Since density of WiFi APs is much higher than cell stations,
\rev{SWFAPRs provide more temporally continuous location data with higher localization accuracy compared with CSRs (usually below 100 meters~\cite{lin2010energy}).}
However, collection of SWFAPRs requires collectors to carry extra devices or install special smartphone APPs), so it is labor-intensive and not very scalable (see example participant size in Table~\ref{table:literature_review}).
Compared with GPS, SWFAPRs have lower localization accuracy,
\rev{but a WiFi scan only consumes around 15 percent of the energy consumed by a GPS location fix~\cite{lin2010energy}}.
\rev{According to Table~\ref{table:literature_review}, SWFAPRs can be applied to four mobility subjects: S.3-6.}

\textbf{GPS Locations (GPSLs):}
\rev{Among all the crowdsensed data types, GPSLs have the best localization performance (usually below 20 meters~\cite{lin2010energy}).}
Similar to SWFAPRs, collection of GPSLs requires extra devices (or smartphone APPs) and hiring collectors. \rev{Therefore, GPSLs datasets have smaller volume and demographic diversity compared with CSRs (hundreds to thousands in magnitude as shown in Table~\ref{table:literature_review}). Compared with SWFAPRs, GPS trackers consume higher energy in mobile devices (around seven times)~\cite{lin2010energy}.}
\rev{According to Table~\ref{table:literature_review}, GPSLs can be applied to four mobility subjects: S.1-4.}

\textbf{Geotagged Social Media (GTSM):} Example GTSM data sources are Twitter and Foursquare. A geotagged tweet contains user identifier, text, and a location (latitude and longitude). 
Publicly available Foursquare data are the venue-side data including location, type, and check-in (visiting) statistics of recorded public venues.
\rev{Similar to CSRs, GTSM have advantages of easy access and large volume
(thousands in magnitude as shown in Table~\ref{table:literature_review}),
and disadvantages of temporal discontinuity and spatial gapping.} In addition, GTSM data can provide more information about location and movement through the textual input from users.
\rev{According to Table~\ref{table:literature_review}, GTSM can be applied to three mobility subjects: S.1-3.}

\textbf{Public Transport Smart Card Records (PTSCRs):} PTSCRs are collected by automated fare collection systems of public transportation services which generally cover bus routes and metro lines. Although PTSCRs data are geographically sparse, incomplete, and biased to people commuting by public transportation, they are useful for gaining insights about the operation of public transportation systems.
\rev{Volumes of PTSCRs datasets are generally very large (millions in magnitude as shown in Table~\ref{table:literature_review}).}
\rev{According to Table~\ref{table:literature_review}, PTSCRs can be applied to three mobility subjects: S.1-3.}

\textbf{Bluetooth Detection Records (BDRs):} Bluetooth detection is a passive data collection technique (no collectors' engagement is needed) which senses surrounding Bluetooth-enabled devices. Unique Bluetooth MAC identifier allows tracking of a particular device.
\rev{Since the typical range of Bluetooth in smartphones is around ten meters~\cite{schauer2014estimating}, BDRs are mainly useful to study building-wide human mobility (see Table~\ref{table:literature_review}).}
BDRs have two main drawbacks. Firstly, extra sensors need to be installed in the desired locations, and secondly, only a small portion of smartphones have Bluetooth enabled.
\rev{According to Table~\ref{table:literature_review}, BDRs can be applied to two mobility subjects: S.5-6.}

\textbf{WiFi Probe Request Records (WFPRRs):} WFPRRs are signals sent out by mobile devices to detect surrounding WiFi APs. They contain MAC address of the mobile device, and thus can be used to track devices.
\rev{The communication range of WiFi technology is between 35 and 100 meters~\cite{schauer2014estimating}.}
Therefore, WFPRRs are mainly suitable for building-wide human mobility subjects (see Table~\ref{table:literature_review}). The collection of WFPRRs also needs to place extra scanners, or to manipulate existing APs. However, it is much more likely for people to turn on WiFi than Bluetooth~\cite{schauer2014estimating}.
\rev{According to Table~\ref{table:literature_review}, WFPRRs can be applied to two mobility subjects: S.5-6.}


\subsection{Human Mobility Subjects}

\rev{Here we describe details of the six mobility subjects
and discuss the suitability of each data type for each subject. Examples of each combination of data type and mobility subject can be found in Table~\ref{table:literature_review}.}
\rev{One should select the most suitable data types given specific research goals and resources.
}

\textbf{S.1 - Distance and Duration Distributions:} This subject is aimed at modeling the probabilistic distributions of certain attributes related to human mobility (e.g. traveling distance and duration). Related studies try to fit different models (e.g. L$\acute{e}$vy walks model) to real-world data and explain the reasons behind the models. From Table~\ref{table:literature_review}, four data types have been used to study this subject. Results obtained from CSRs are more representative of the population, but the studied attributes are limited to mobility flights (i.e. jumps from origins to destinations). GPSLs can yield more fine-grained mobility attributes and is more flexible in terms of selecting sample group. GTSM are more suitable for studying attributes related to place visiting (e.g. probability of users returning to a given location), but it is hard to clarify the covered population if no personal information is available. PTSCRs are limited to explore trips by public transportation only.

\textbf{S.2 - Origin-Destination (OD) Matrices:} This subject is to develop OD matrices which describe traffic conditions between important nodes in traffic networks. The obtained values can be either people counts or average traveling time. Four data types have been used to develop these matrices. CSRs yield results with wider spatial coverage and more general numbers due to their large volume. But it is difficult to map CSRs to desired traffic network nodes because of the large spatial granularity of cell towers. GPSLs can generate matrices between more precise traffic nodes but have problems of limited data size and heavy manpower. Due to the randomness and large spatial gapping of GTSM data, they are often used to build matrices between large city areas instead of specific traffic nodes. For the nature of WFPRRs, they are only suitable for building matrices between bus stops and metro stations.

\textbf{S.3 - Individual Activity-Based Mobility Patterns:} This subject is about gaining insights of people's activities (visits to places), such as identifying activity categories (e.g. home, working, shopping)~\cite{hasan2013understanding} and predicting next visiting places~\cite{noulas2012mining}. Five data types have been used. CSRs are relatively easy to get, but it is difficult to extract human activities from CSRs due to large spatiotemporal gapping. SWFAPRs and GPSLs have better location precision, and thus are suitable for fine-grained research tasks, such as exploring human contacts~\cite{vazquez2013using}. Their collection, however, is labor-intensive and not very scalable. Due to access to text, it is easier to extract activity information using GTSM but attentions should be paid to sample bias issue. Although PTSCRs can be used, they are limited to simpler tasks such as identifying one's home and working locations.

\textbf{S.4 - Individual Transportation Mode Identification:} This subject is to identify individual transportation modes (e.g. walking, driving, riding buses). Therefore, it requires continuous tracking of one's locations and thus only CSRs, SWFAPRs, and GPSLs have been used for this subject. CSRs have the worst location accuracy and temporal continuity, so they can only train simpler algorithms to identify basic modes (e.g. walking and vehicle). GPSLs have the best localization accuracy and continuity, and thus train better algorithms to identify more detailed modes (e.g. cycling, running, riding buses). SWFAPRs have the medium level of localization accuracy and continuity, so it is feasible yet more difficult to train good algorithms with them, but collecting SWFAPRs consumes much less power than collecting GPSLs.

\textbf{S.5 - Densities and Flows within a Building:} This subject is focused on human density and movement in a large building. Four data types have been used in related studies. CSRs do not require extra sensors and are collected passively, but enough number of cell stations and access to the data are needed. SWFAPRs yield more fine-grained results due to more accurate locations, but their collection is labor-intensive and not very scalable. BDRs and WFPRRs have medium location accuracy and require installation of extra sensors, but their allocations are flexible and collection is scalable. In addition, WFPRRs sense more unique devices compared with BDRs~\cite{schauer2014estimating}.

\textbf{S.6 - Densities and Flows within a Cluster of Buildings:} This subject is to understand people's density and movement between near-by buildings (e.g. hospital complex, commercial district, university campus). SWFAPRs, BDRs, and WFPRRs have been used in related studies. Their pros and cons to this subject are same as those to the above subject.


\section{Data Preprocessing and Analysis Methods}


\subsection{Data Preprocessing}

\rev{
Data preprocessing includes data transforming and noise handling. The former ensures suitable data format for subsequent analysis, while the latter is necessary for good results. 

Noises in crowdsensed mobility data can be classified as system noises and human noises. System noises stand for bias and errors caused by imperfection of data collection systems which mainly exist in CSRs, SWFAPRs, GPSLs, BDRs, and WFPRRs. They are normally handled by filtering and interpolation.
Human noises are bias and errors introduced by human behaviors, mainly existing in GTSM. Because social media is used for communication purposes and the geographic tags are added by users, the collected data can be highly affected by users' moods, thoughts, and outside incentives~\cite{rost2013representation}. Therefore, careless usage of GTSM for human mobility studies causes large discrepancies between the results and actual mobility~\cite{wang2016will}. Due to the lack of relative information in GTSM datasets, it is difficult to handle the human noise.}

\subsection{Common Data Analysis Methods}

\rev{Here, seven common data analysis methods are introduced with their applications to understanding urban human mobility.}

\rev{\textbf{M.1 - Visualization:} Data visualization is a powerful tool for understanding mobility data and presenting patterns. As shown in Table~\ref{table:literature_review}, it is indispensable in mobility research. Common visualization formats are summarized as follows.
A scatter plot on a map is used to visualize trajectories and important locations, such as the trip example in~\cite{ghorpade2015integrated} and base station map in~\cite{zhong2017characterizing}.
A bar chart or histogram is often used to study mobility attribute values that are accumulated in consecutive time intervals, such as the histogram of flow densities in~\cite{demissie2016inferring}.
A heatmap can reveal knowledge about spatial or temporal distributions of certain type of human activities, such as the spatial distribution of user counts in~\cite{jiang2017activity} and temporal distribution of journey time in~\cite{mohamed2017clustering}.}

\rev{\textbf{M.2 - Statistics:} Statistical analysis used for human mobility research mainly include three types:
\begin {enumerate*}
\item extracting conclusive statistics from complex mobility attributes or research results, such as classification evaluation matrices presented in~\cite{ghorpade2015integrated}, 
\item fitting statistical models to describe mobility attributes, such as fitting truncated power laws to social media check-in data in~\cite{hasan2013understanding},
\item conducting statistical test to evaluate research findings, such as using Monte Carlo simulations to test significance of wavelet analysis in~\cite{vazquez2013using}.
\end {enumerate*}}

\textbf{M.3 - Heuristic Logic:} Heuristic logic means setting data processing rules (e.g. thresholds) based on life experience or common logic. It is often used when detecting staying spots from raw location data or identifying places of interest (PoIs) from staying spots.
\rev{For example, in~\cite{jiang2017activity}, to extract staying spots by clustering call detail record data, they preset the distance threshold as 300 meters and time threshold as 10 minutes.}

\textbf{M.4 - Graph Theory:} This method builds graph-based model from data to investigate connections between people's staying (or visited) spots. For example, daily motifs are extracted in~\cite{jiang2017activity} to describe the daily patterns of people's movement between staying spots. In~\cite{vazquez2013using}, complex network model is built to investigate connections between locations visited by the same individuals.

\rev{\textbf{M.5 - Optimization:} Optimization techniques are usually adopted when there is a need to fit statistical models or to solve context-fitting problems. In~\cite{vazquez2013using}, maximum likelihood methods are applied to fit distributions to distance values. Context-fitting problems include mapping trips to origin-destination pairs~\cite{ge2016updating} and estimating the alighting locations of smartcard data based on boarding information~\cite{munizaga2012estimation}.}

\textbf{M.6 - Clustering:} In mobility research, clustering is mainly used to:
\begin {enumerate*}
\item find people's staying spots and PoIs~\cite{jiang2017activity}, 
\item cluster people into different groups by their mobility patterns~\cite{mohamed2017clustering}, 
\item cluster cell stations (or BDRs and WFPRRs sensing stations) into different groups by their utility patterns~\cite{zhong2017characterizing}. 
\end {enumerate*}
Popular algorithms used are Density-based spatial clustering of applications with noise (DBSCAN) and k-means clustering.

\textbf{M.7 - Classification:} In mobility research, classification is mainly used to:
\begin {enumerate*}
\item identify the categories of places people visit~\cite{hasan2013understanding}, 
\item classify people into different profiles~\cite{zhong2017characterizing}, and 
\item identify people's transportation modes~\cite{ghorpade2015integrated}. 
\end {enumerate*}
Usually, many classifiers (e.g. tree-based models, support vector machine, artificial neural network) can be applied, and a trial-and-error process is often necessary to select the best one.


\section{City-Wide Case Study}

In this case study, we present our research project, the national science experiment (NSE) of Singapore, which collects and analyzes large-scale SWFAPRs data to understand city-wide young students' mobility.
\revtwo{Compared with previous studies, this work has a much larger-scale data collection and obtains precise individual activity and transportation usage patterns from only SWFAPRs data.}

\begin{figure}[!t]
\centering
\includegraphics[width=\linewidth]{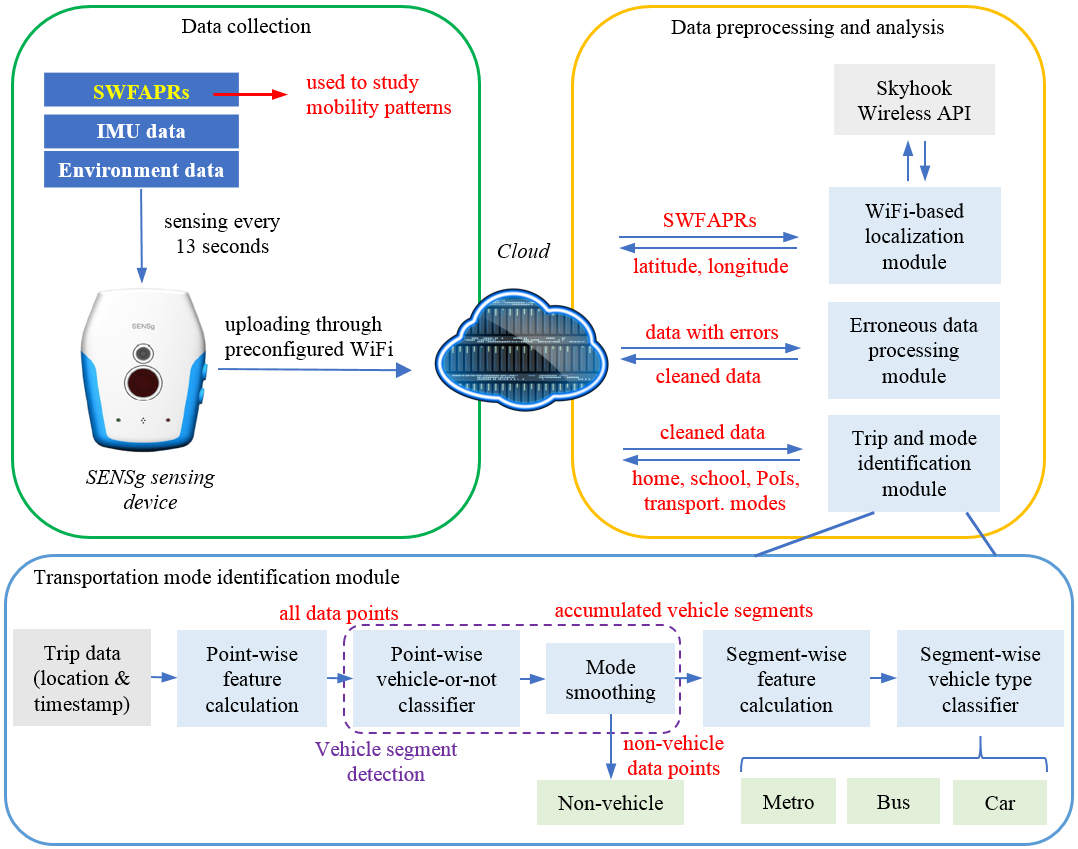}
\caption{Data pipeline of the NSE project.}
\label{fig_data_collection_preproseccing}
\end{figure}

\subsection{Data Collection, Preprocessing, and Labeling}

Data collectors of this study are students from primary, secondary and high schools. To avoid affecting students' normal life, we designed a sensing device (called SENSg) which can work consecutively for one week without charging and collect different types of data (see Fig.~\ref{fig_data_collection_preproseccing}). To restrict power consumption, SWFAPRs are collected instead of GPSLs as the location data source.
\rev{Although inertia measurement unit (IMU) and environment data are collected, they are used for other studies.
Only SWFAPRs are used to study students' mobility patterns here.}
In 2015 and 2016, we had 90,991 participants from over 100 schools spread over the city.

During data preprocessing (see Fig.~\ref{fig_data_collection_preproseccing}), WiFi-based localization is conducted first to map SWFAPRs to location values (latitude and longitude) by calling API of Skyhook Wireless. 
Afterwards, erroneous data are processed by filtering out data points with abnormal sensor values (including location values).

\rev{
Since we want to identify students' transportation modes using classification, ground truth labels are necessary. But it is impractical to ask young students to provide good-quality data labels. To make use of NSE dataset, we labeled the data by ourselves. In general, traveling between same OD pair using different transportations yields different geographic trajectories, distance, and duration. By comparing these features between the collected trip and all possible routes between the same OD (available from Google Maps Directions API), one can often find a very similar route to the collected trip. Since the transportation modes of that similar route are known, the modes of the collected trip can be labeled accordingly.}

\subsection{Data Analysis and Results}

Data analysis is conducted step by step as follows.

\textbf{Extracting staying spots:} Cleaned data are first processed to extract students' staying spots. When the moving speed (calculated from location values) is below one preset threshold for a certain time period, a staying behavior is captured. Location values of all points belonging to this stay are averaged to obtain the location of the staying spot.

\begin{figure}[!b]
\centering
\subfloat[]{\includegraphics[width=.85\linewidth]{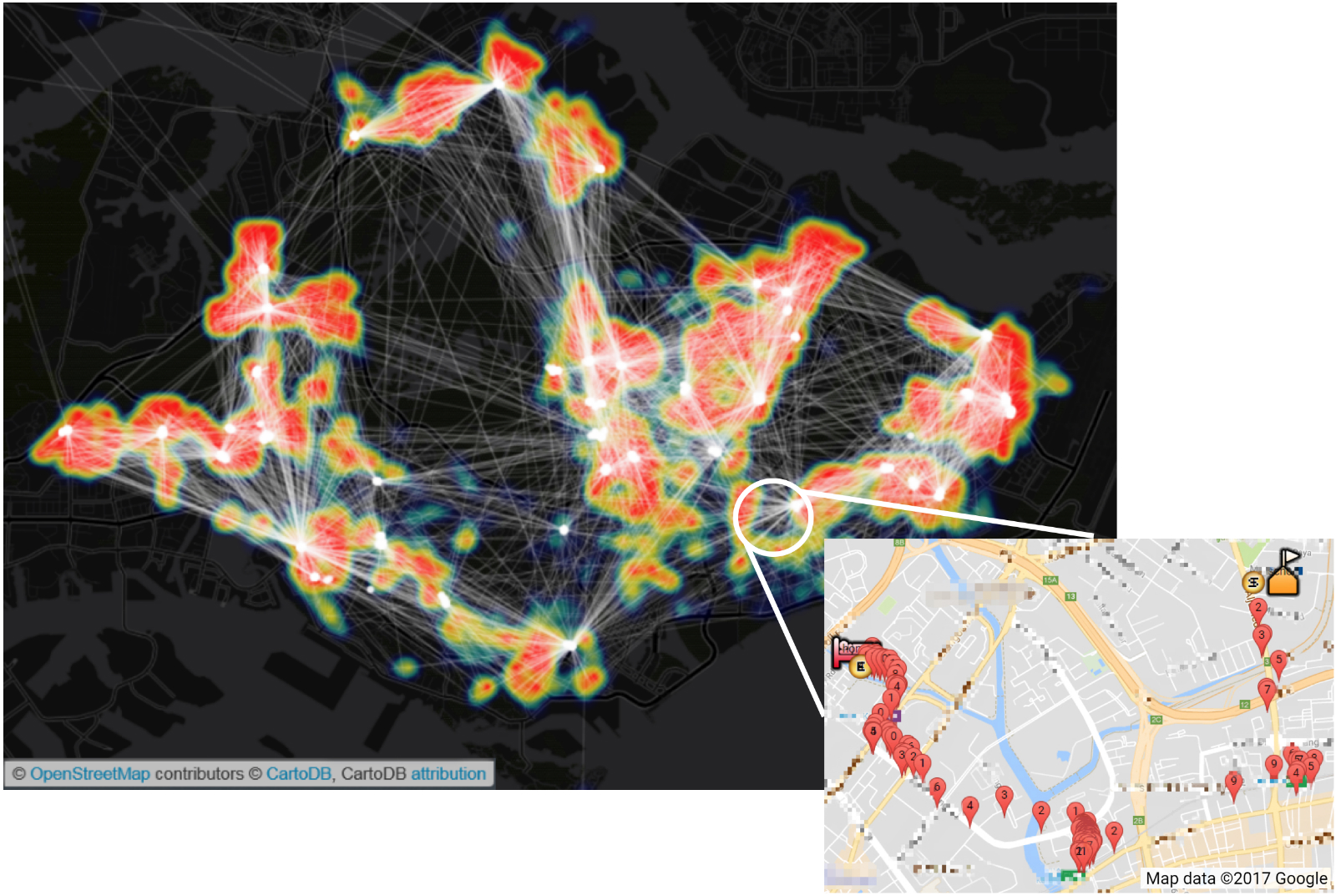}%
\label{fig_home_school}}
\vfill
\subfloat[]{\includegraphics[width=.9\linewidth]{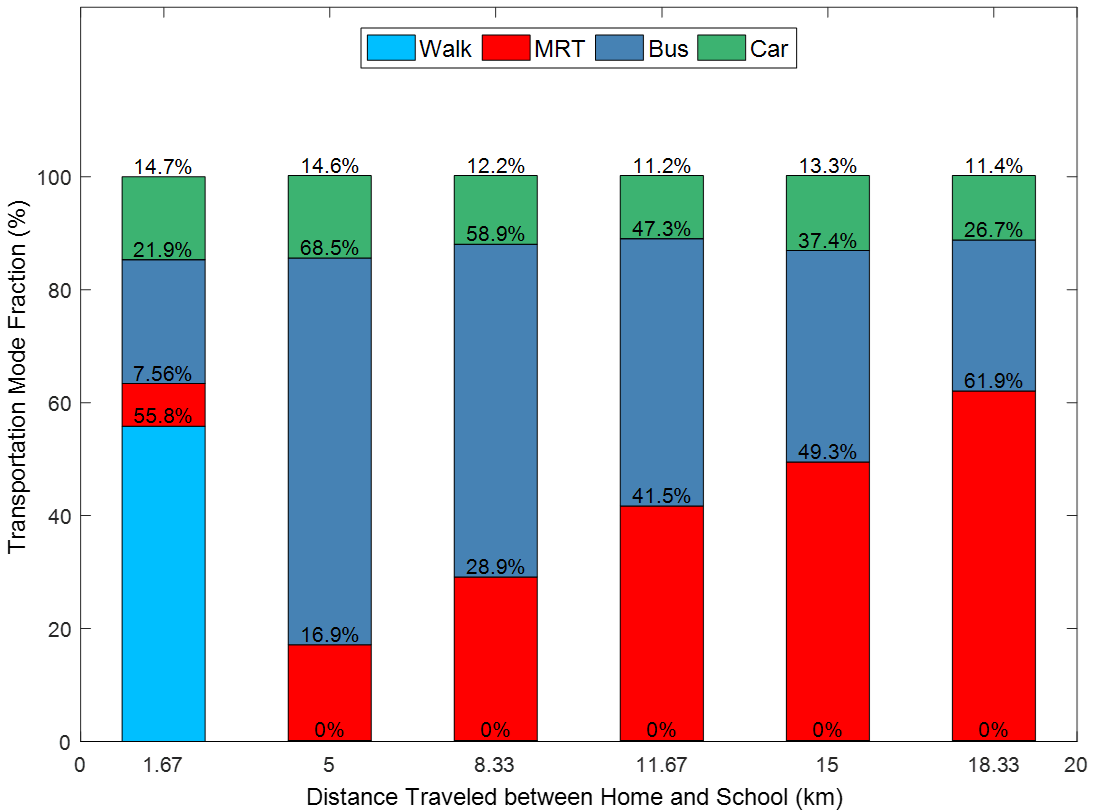}%
\label{fig_mode_frac}}
\caption{\textbf{NSE data analysis results}: 
\rev{(a)~city-wide heatmap of students' home distribution with links to schools (white dots) and an example trip trajectory, (b)~changing trend of students' daily transportation mode fractions with the increase of traveled distance between home and school (the counting unit is individual trip captured in the second analysis step).}}
\label{fig_NSE_results}
\end{figure}

\textbf{Identifying PoIs and trips:} PoIs represent functional places where students stay (e.g. shopping malls and schools). Since one can still move in a large building, PoI represents a wider area than staying spots. DBSCAN is applied here to cluster the extracted staying spots to detect PoIs. Data points between each chronological pair of PoIs therefore form up a trip. \rev{Figure~\ref{fig_home_school} shows an example trip obtained in this step.}

\textbf{Identifying home and school:} Home and school are identified from the PoIs by checking their starting and ending time. If the time period is long and overnight, the corresponding PoI is identified as home. If the time period overlaps with the preset school time largely, that PoI is marked as school.

\begin{figure*}[!t]
	\centering
	\begin{tabular}{@{}cccccccc@{}}
		\multirow{4}{*}{\rotatebox{90}{\hspace{0.6cm} Building $G$ \hspace{0.4cm} Building $F$ \hspace{0.4cm} Building $A$\hspace{0.45cm}}}
		& {\fontsize{8}{8}\selectfont (a) Monday } & {\fontsize{8}{8}\selectfont (b) Tuesday } & {\fontsize{8}{8}\selectfont (c) Wednesday } & {\fontsize{8}{8}\selectfont (d) Thursday } &{\fontsize{8}{8}\selectfont (e) Friday } & {\fontsize{8}{8}\selectfont (f) Saturday } & {\fontsize{8}{8}\selectfont (g) Sunday }\\
		&
		\includegraphics[trim=0.6cm 0.0cm 0.7cm 0.0cm ,width=0.134\textwidth]{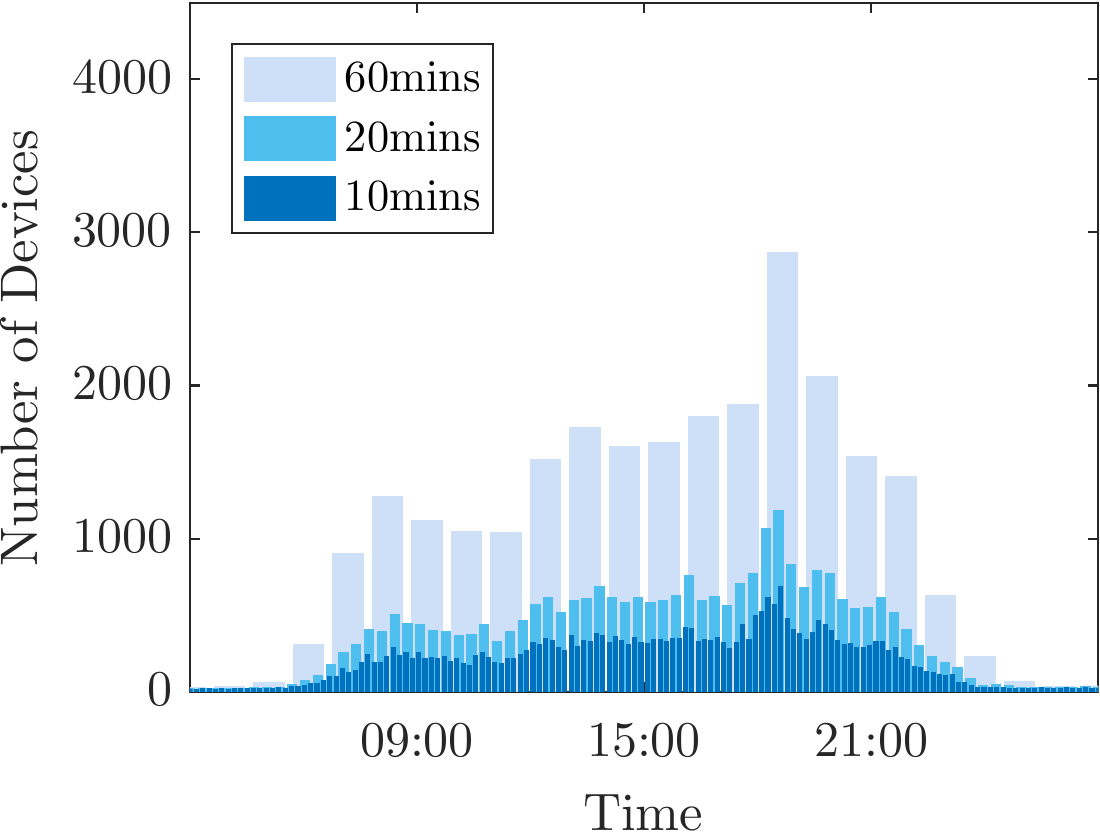}& 
		\includegraphics[trim=0.9cm 0.0cm 0.7cm 0.0cm ,width=0.11\textwidth]{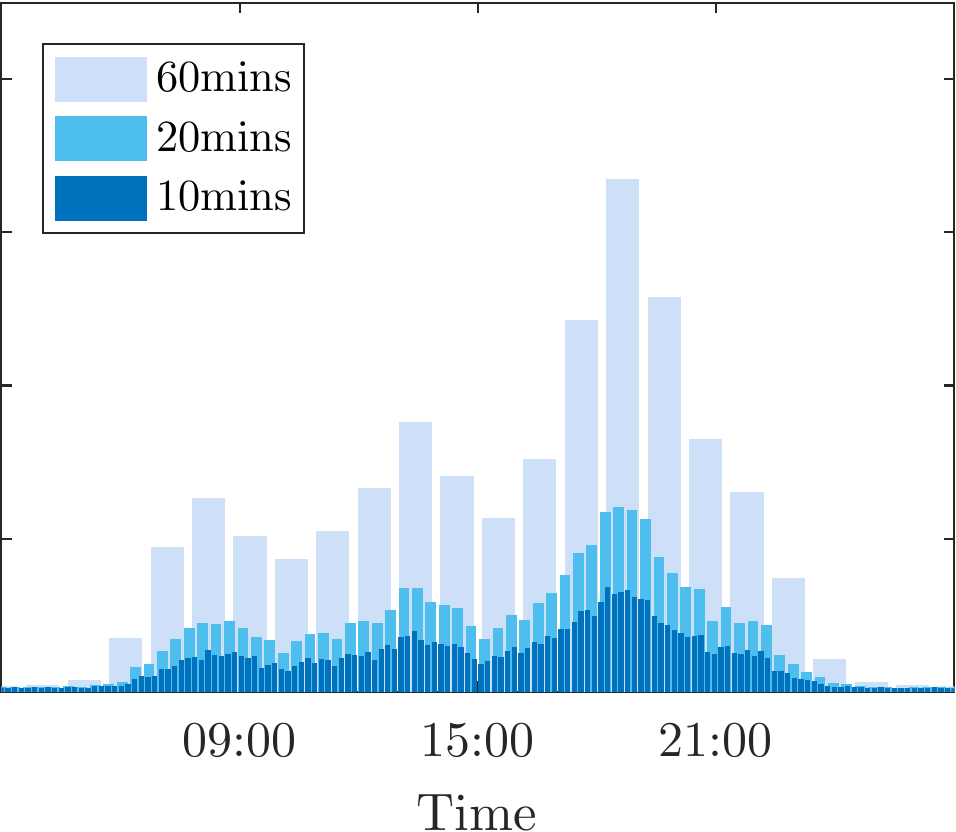}&  
		\includegraphics[trim=0.9cm 0.0cm 0.7cm 0.0cm ,width=0.11\textwidth]{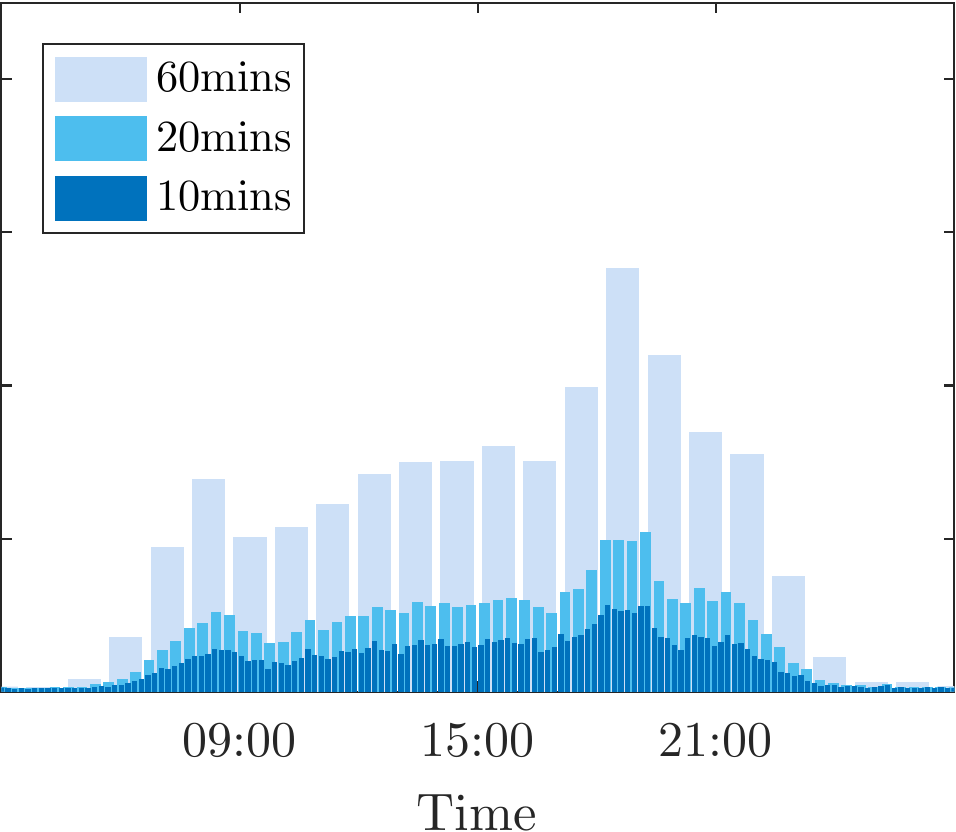}&
		\includegraphics[trim=0.9cm 0.0cm 0.7cm 0.0cm ,width=0.11\textwidth]{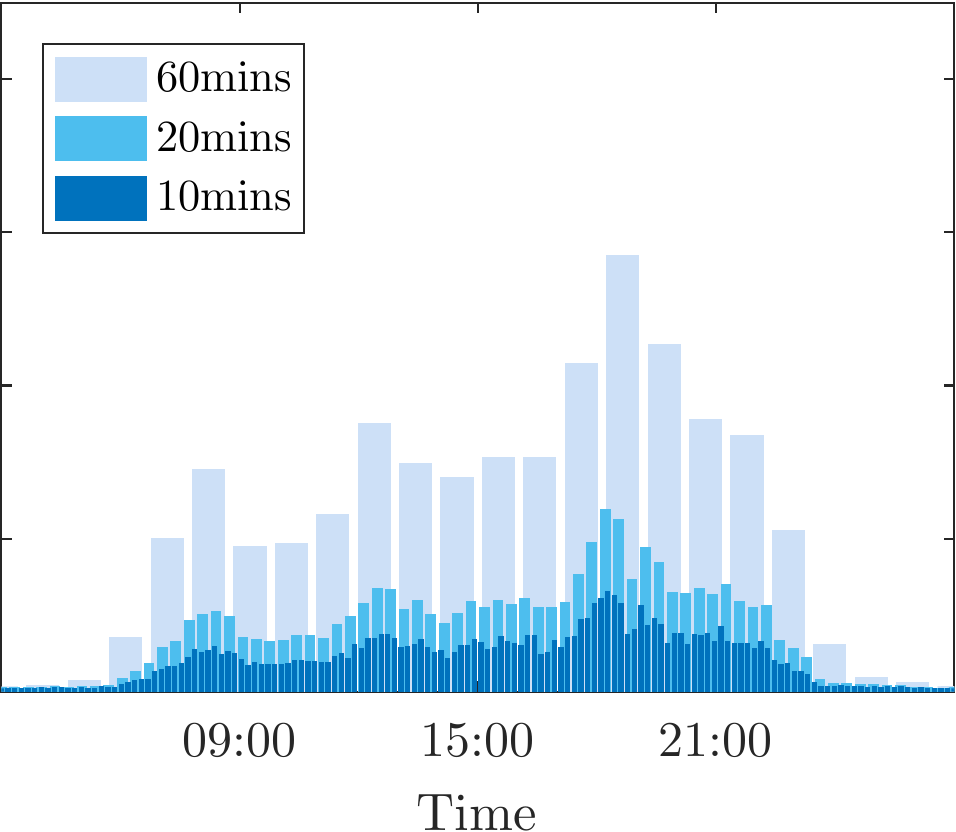}& 
		\includegraphics[trim=0.9cm 0.0cm 0.7cm 0.0cm ,width=0.11\textwidth]{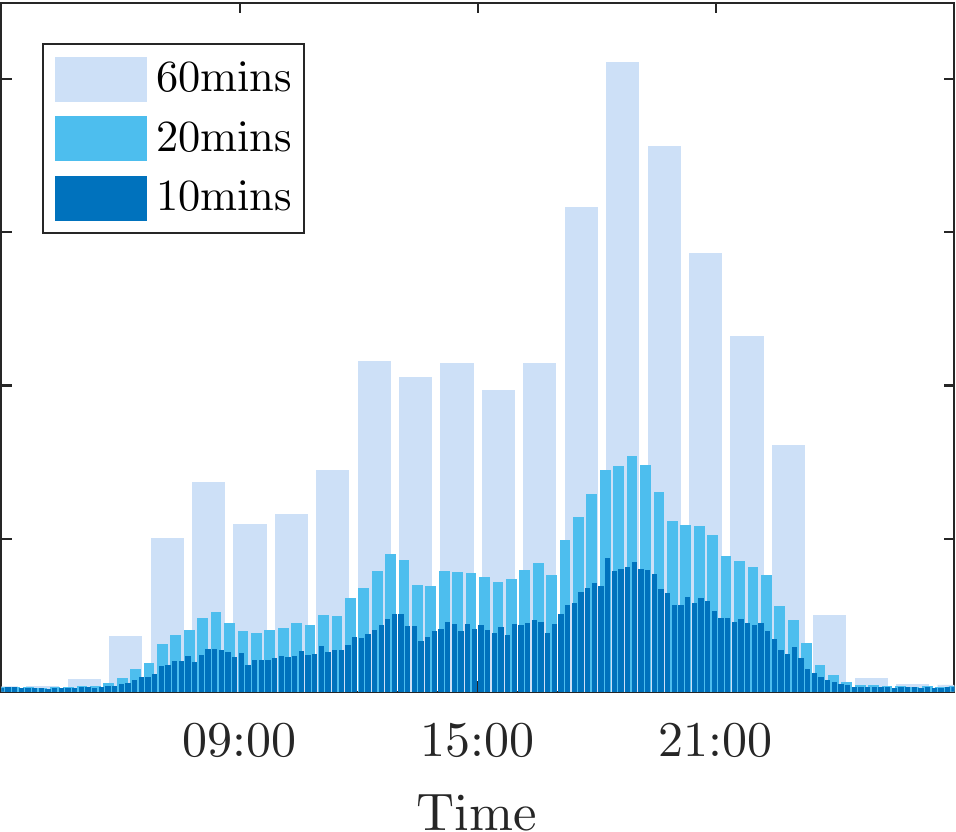}&  
		\includegraphics[trim=0.9cm 0.0cm 0.7cm 0.0cm ,width=0.11\textwidth]{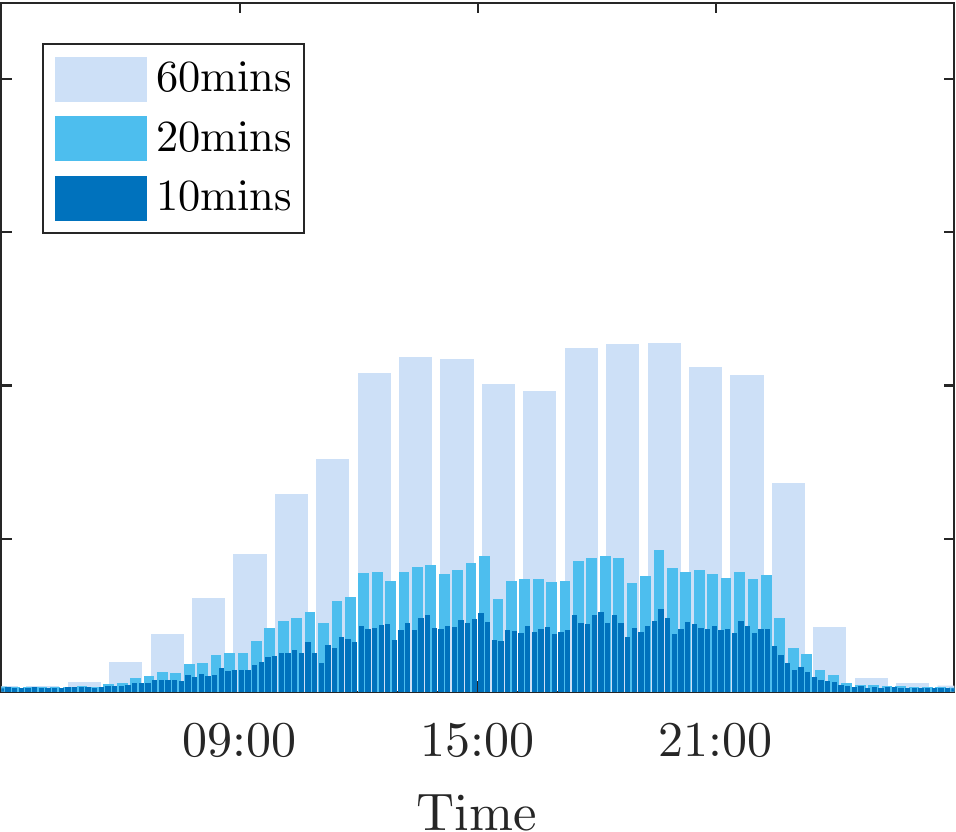} &  
		\includegraphics[trim=0.9cm 0.0cm 0.7cm 0.0cm ,width=0.11\textwidth]{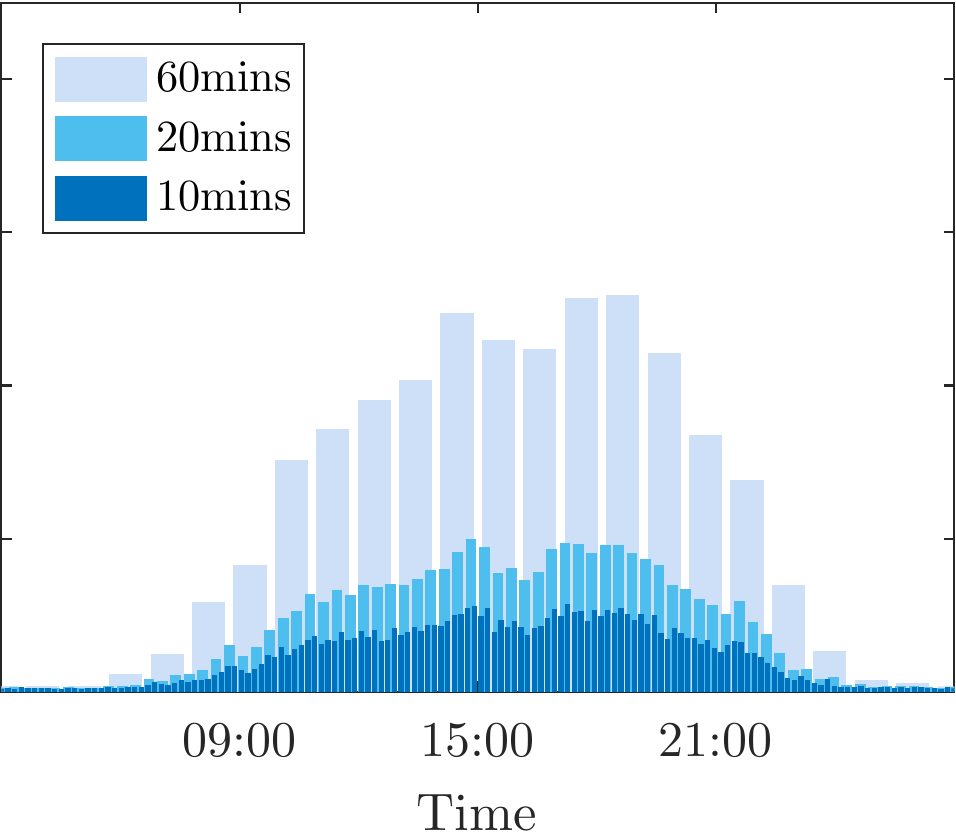} \\
		&
		\includegraphics[trim=0.6cm 0.0cm 0.7cm 0.0cm ,width=0.134\textwidth]{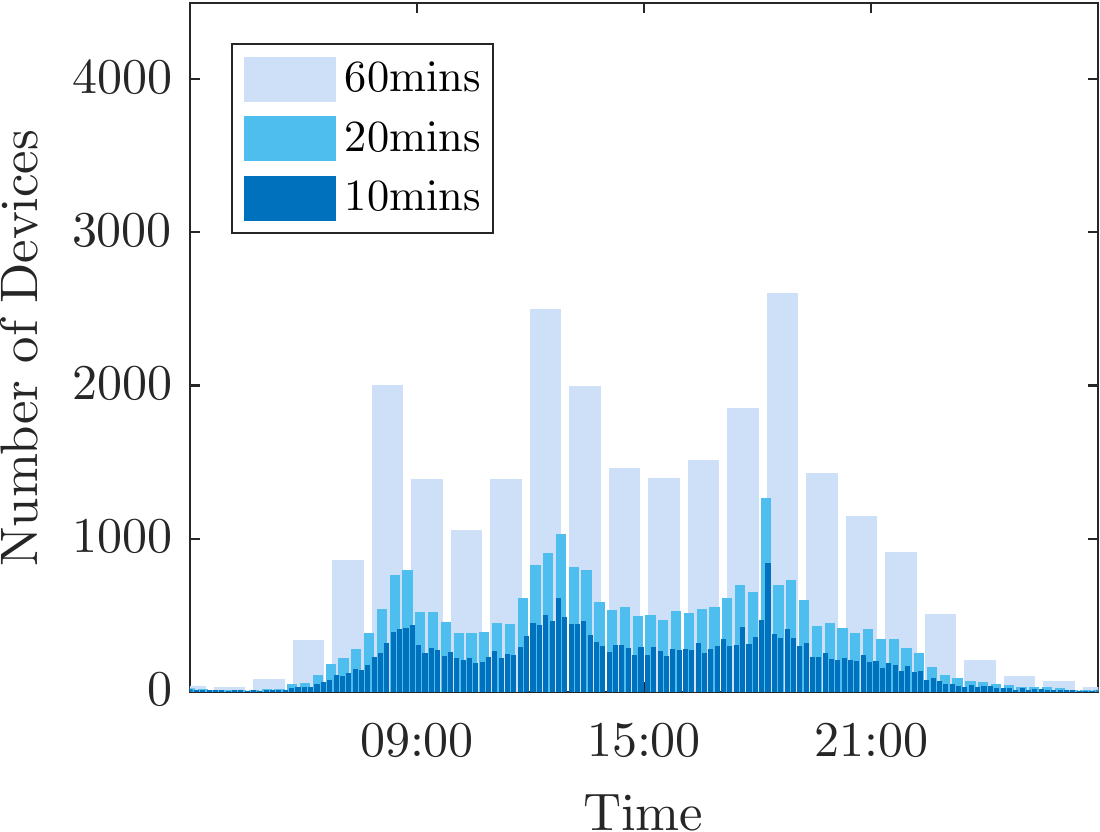}& 
		\includegraphics[trim=0.9cm 0.0cm 0.7cm 0.0cm ,width=0.11\textwidth]{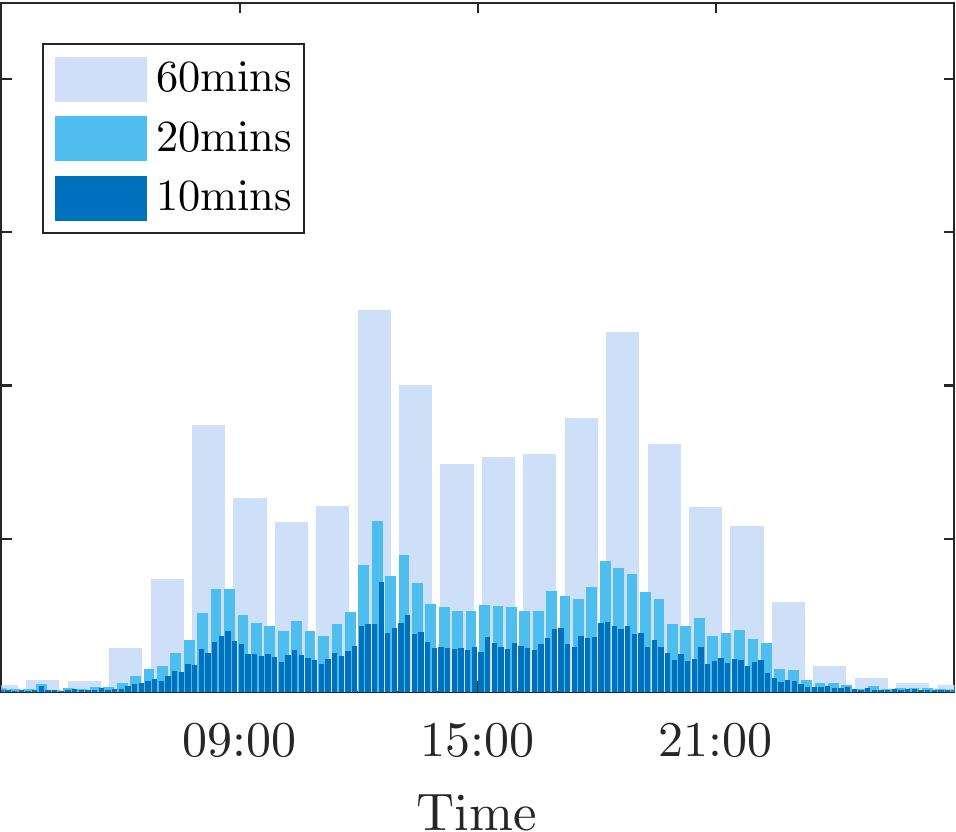}&  
		\includegraphics[trim=0.9cm 0.0cm 0.7cm 0.0cm ,width=0.11\textwidth]{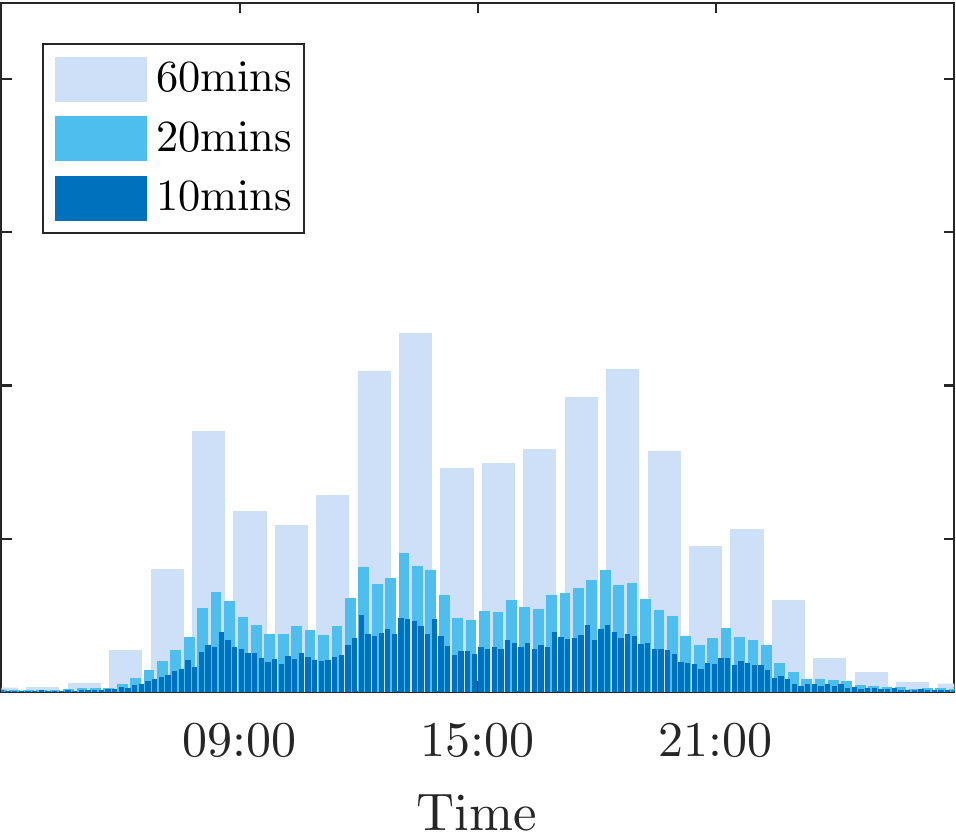}&
		\includegraphics[trim=0.9cm 0.0cm 0.7cm 0.0cm ,width=0.11\textwidth]{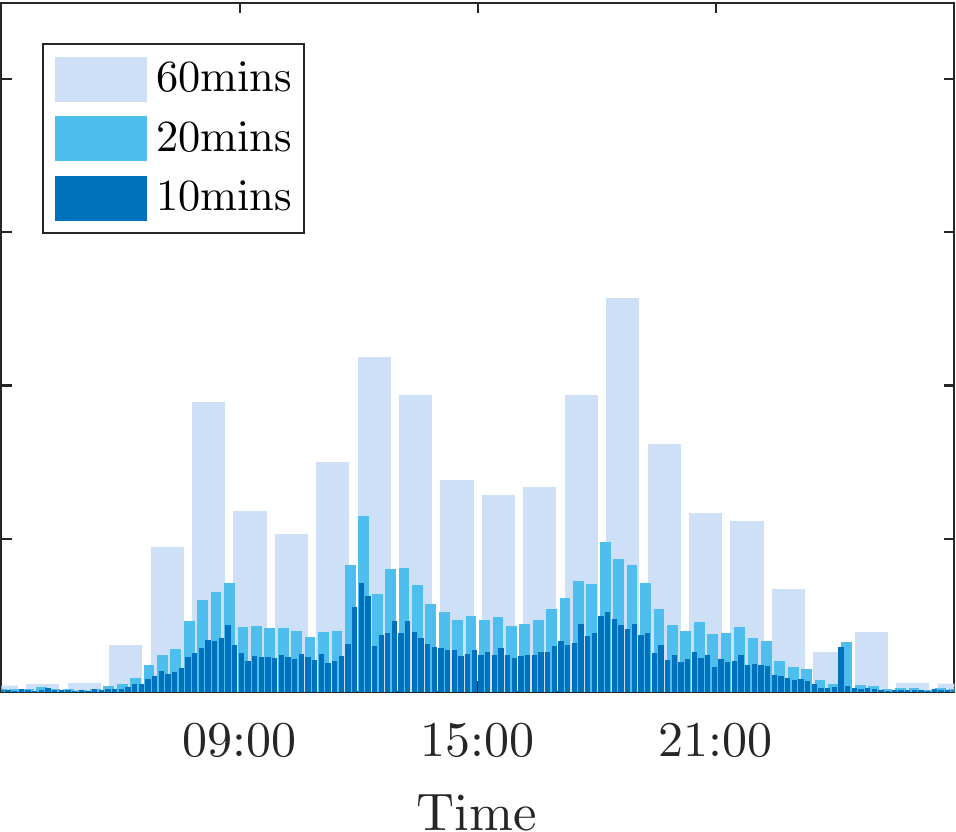}& 
		\includegraphics[trim=0.9cm 0.0cm 0.7cm 0.0cm ,width=0.11\textwidth]{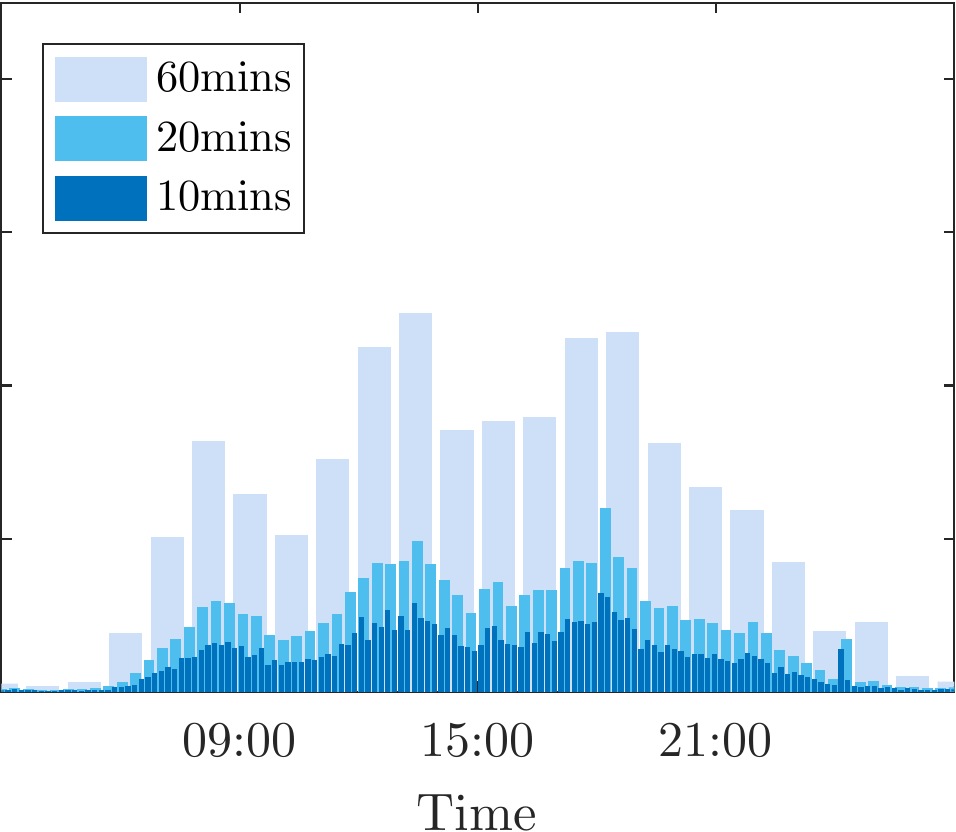}&  
		\includegraphics[trim=0.9cm 0.0cm 0.7cm 0.0cm ,width=0.11\textwidth]{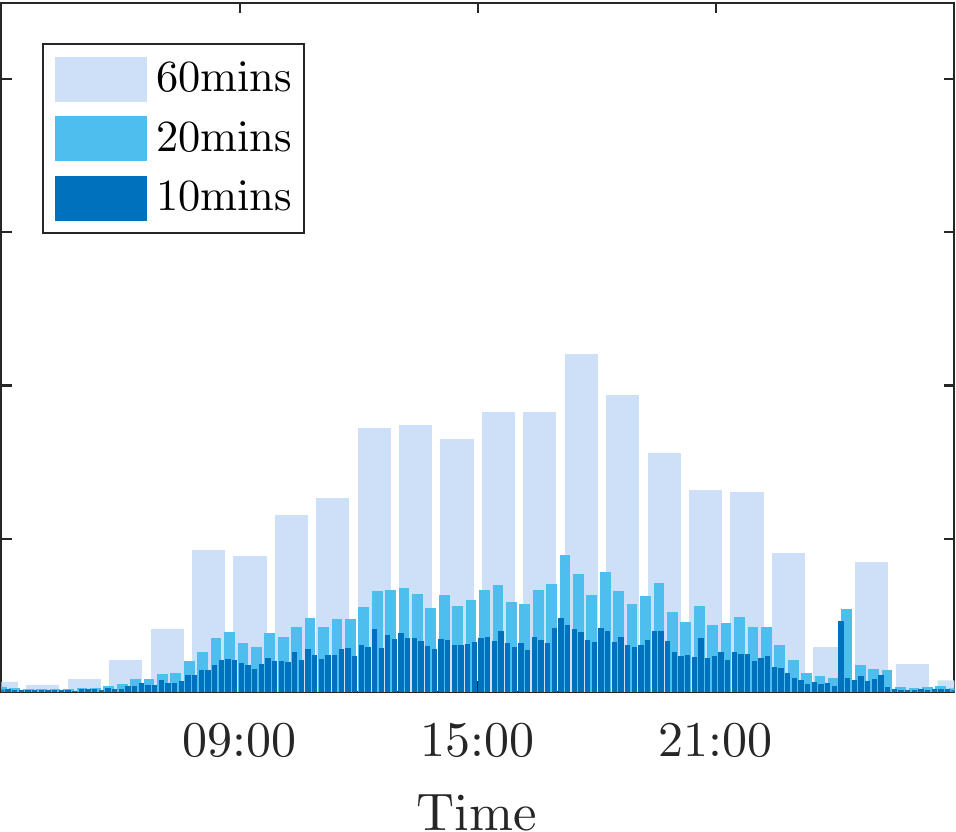} &  
		\includegraphics[trim=0.9cm 0.0cm 0.7cm 0.0cm ,width=0.11\textwidth]{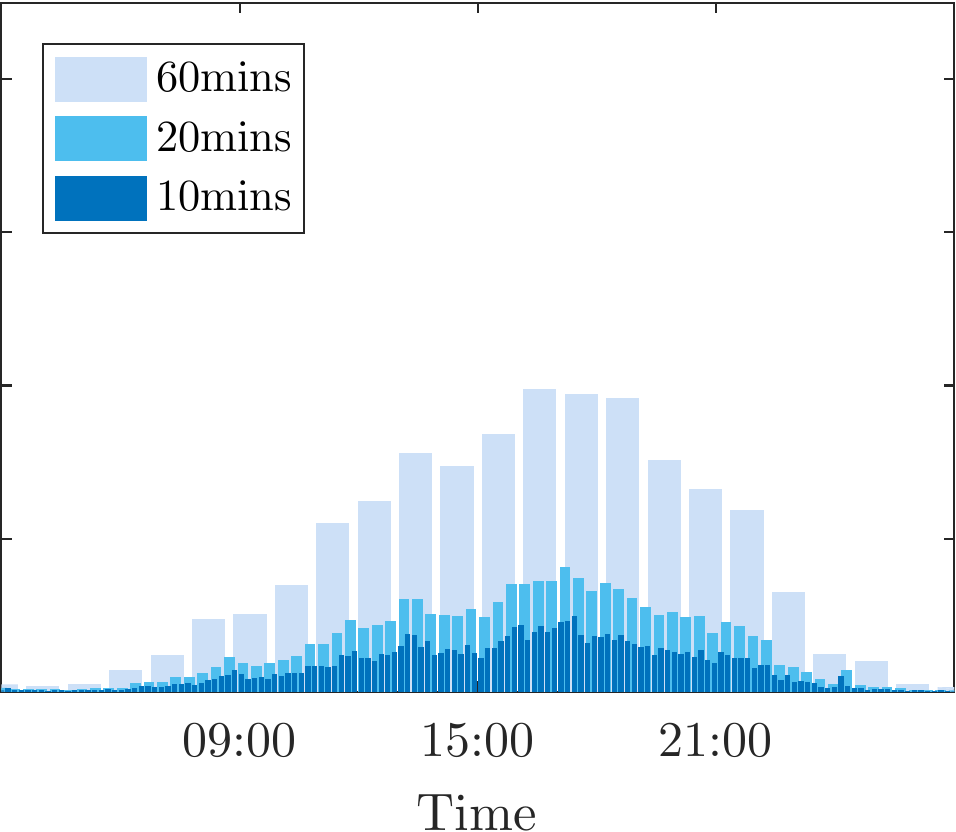} \\
		&
		\includegraphics[trim=0.6cm 0.0cm 0.7cm 0.0cm ,width=0.134\textwidth]{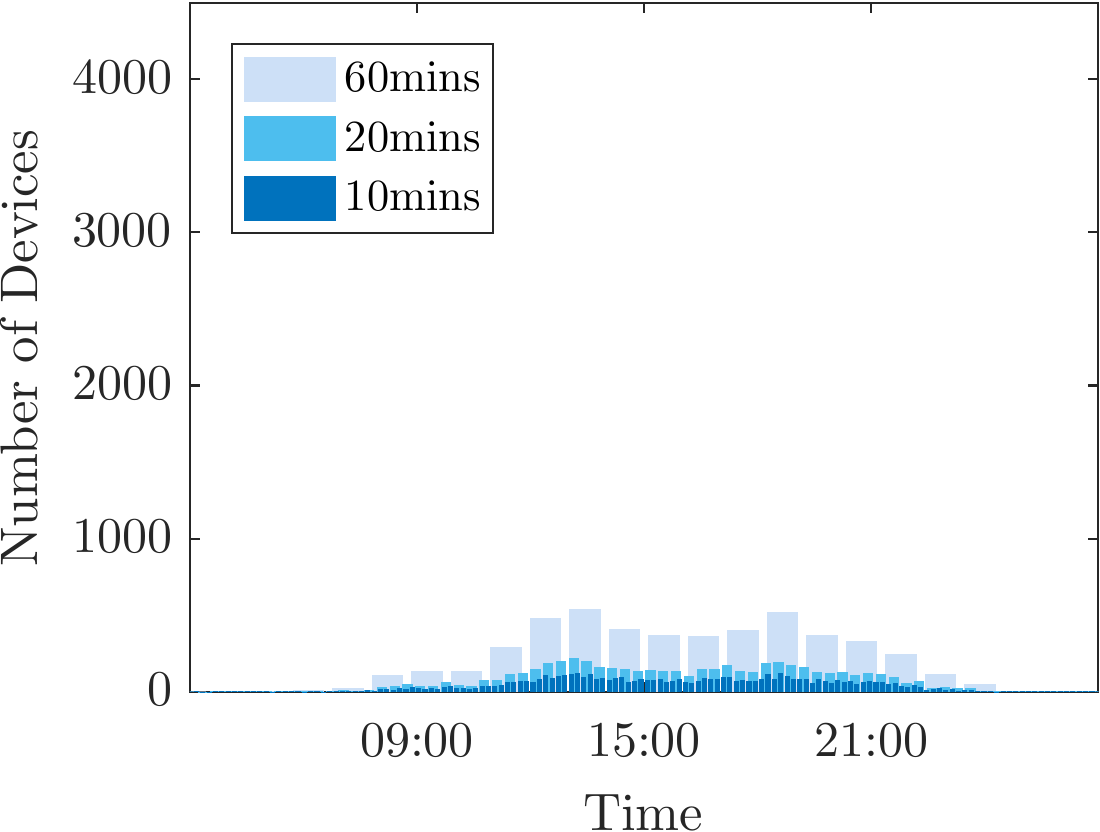}& 
		\includegraphics[trim=0.9cm 0.0cm 0.7cm 0.0cm ,width=0.11\textwidth]{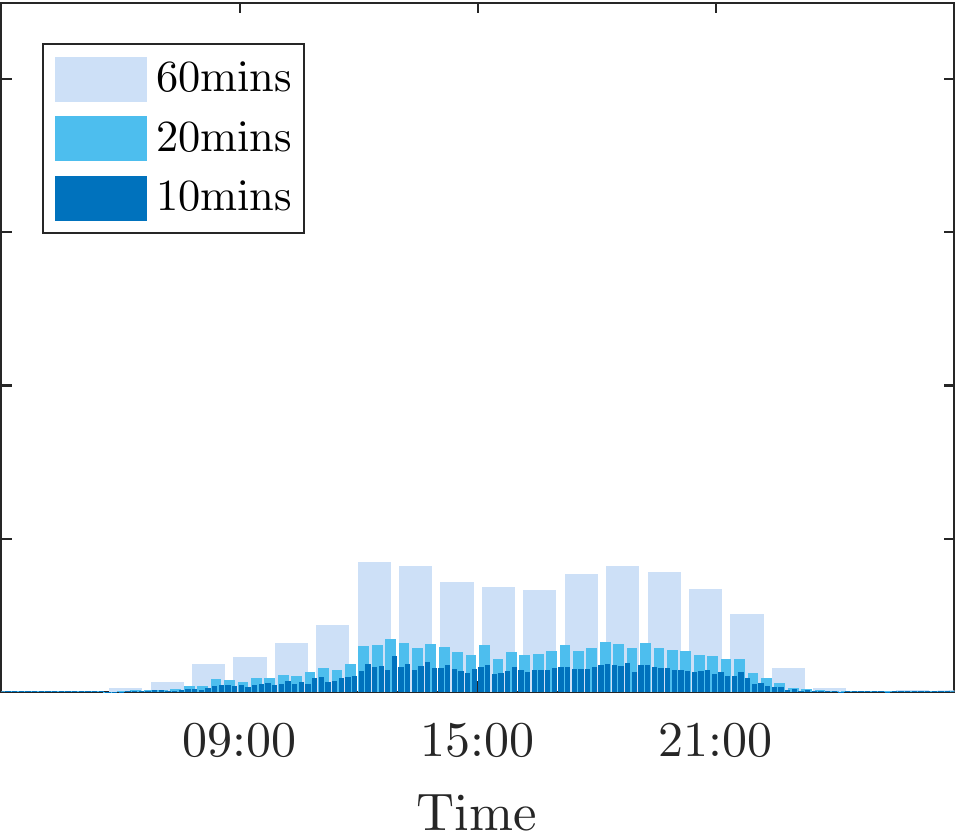}&  
		\includegraphics[trim=0.9cm 0.0cm 0.7cm 0.0cm ,width=0.11\textwidth]{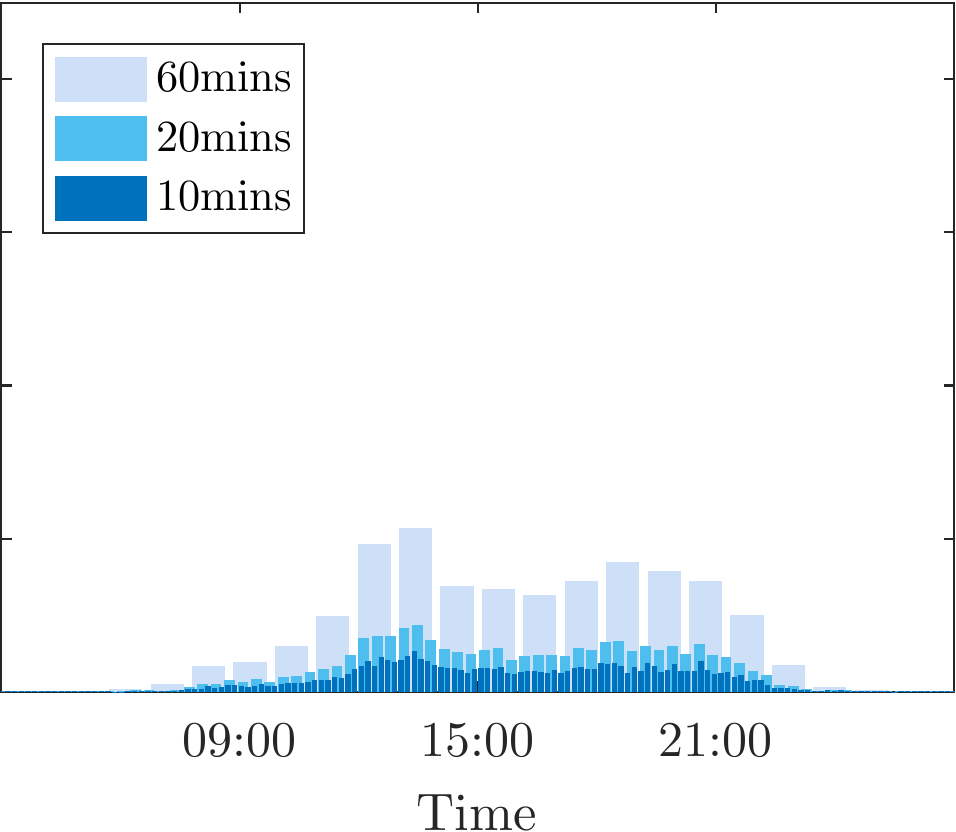}&
		\includegraphics[trim=0.9cm 0.0cm 0.7cm 0.0cm ,width=0.11\textwidth]{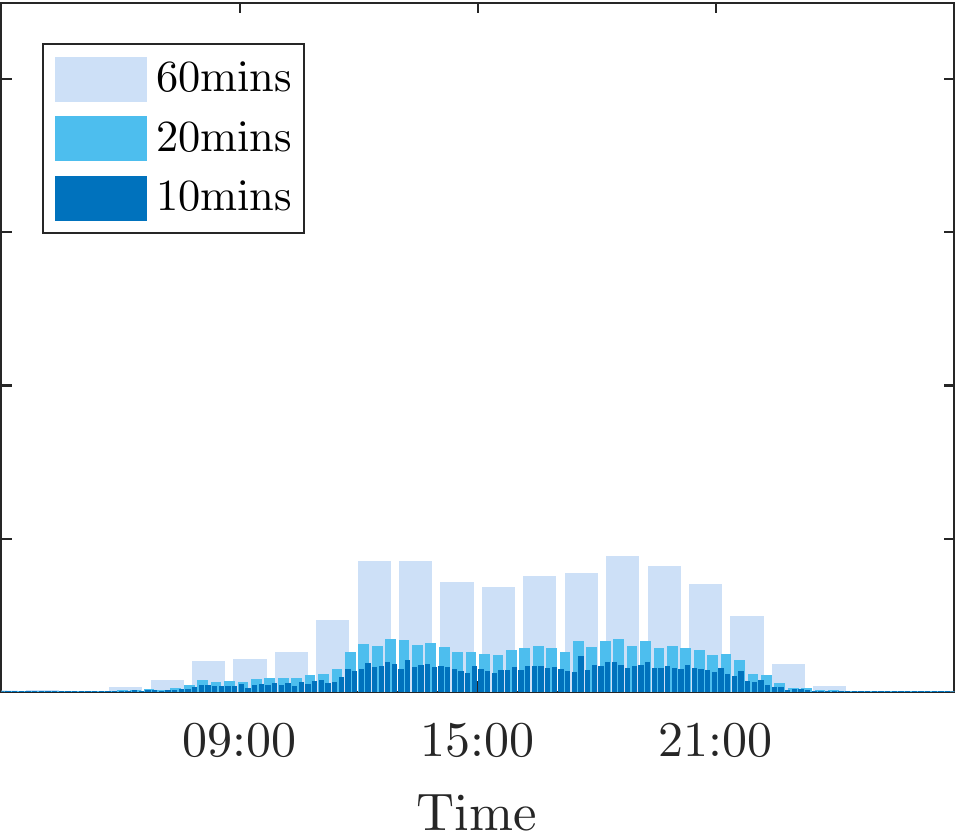}& 
		\includegraphics[trim=0.9cm 0.0cm 0.7cm 0.0cm ,width=0.11\textwidth]{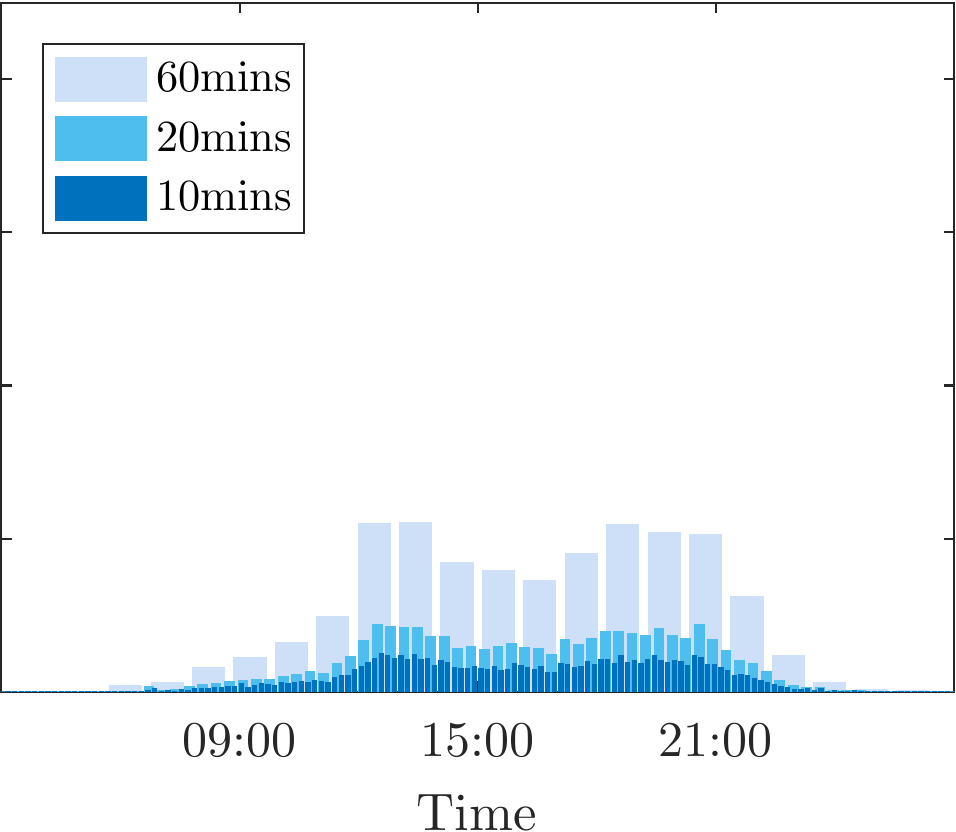}&  
		\includegraphics[trim=0.9cm 0.0cm 0.7cm 0.0cm ,width=0.11\textwidth]{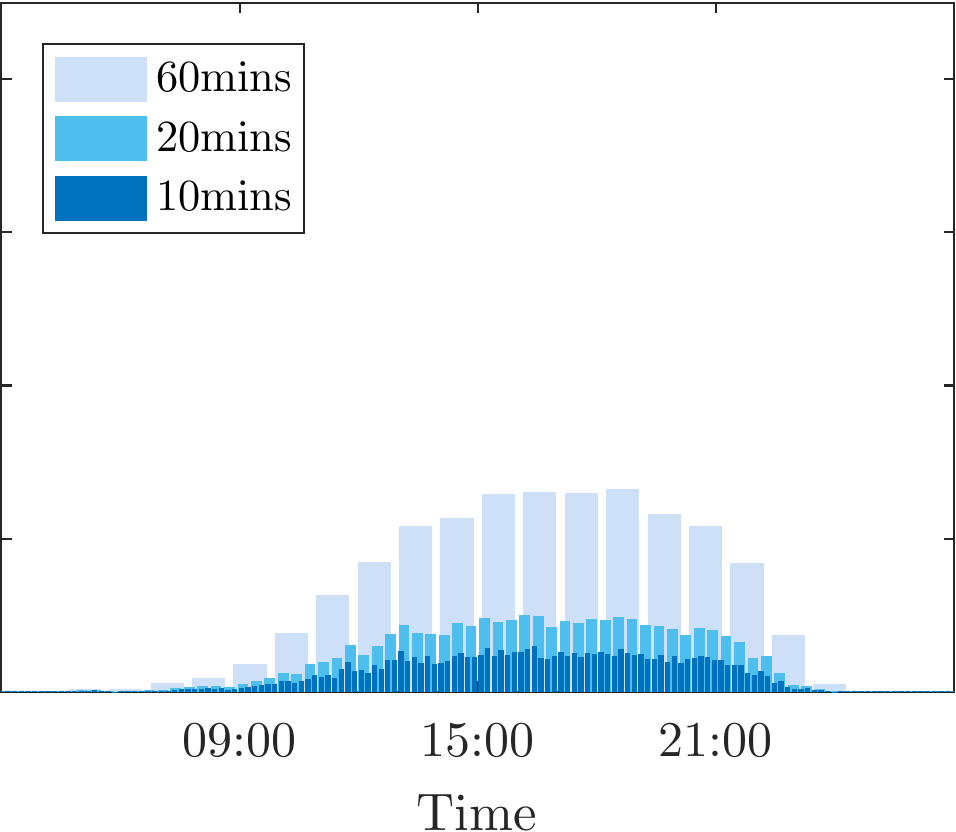} &  
		\includegraphics[trim=0.9cm 0.0cm 0.7cm 0.0cm ,width=0.11\textwidth]{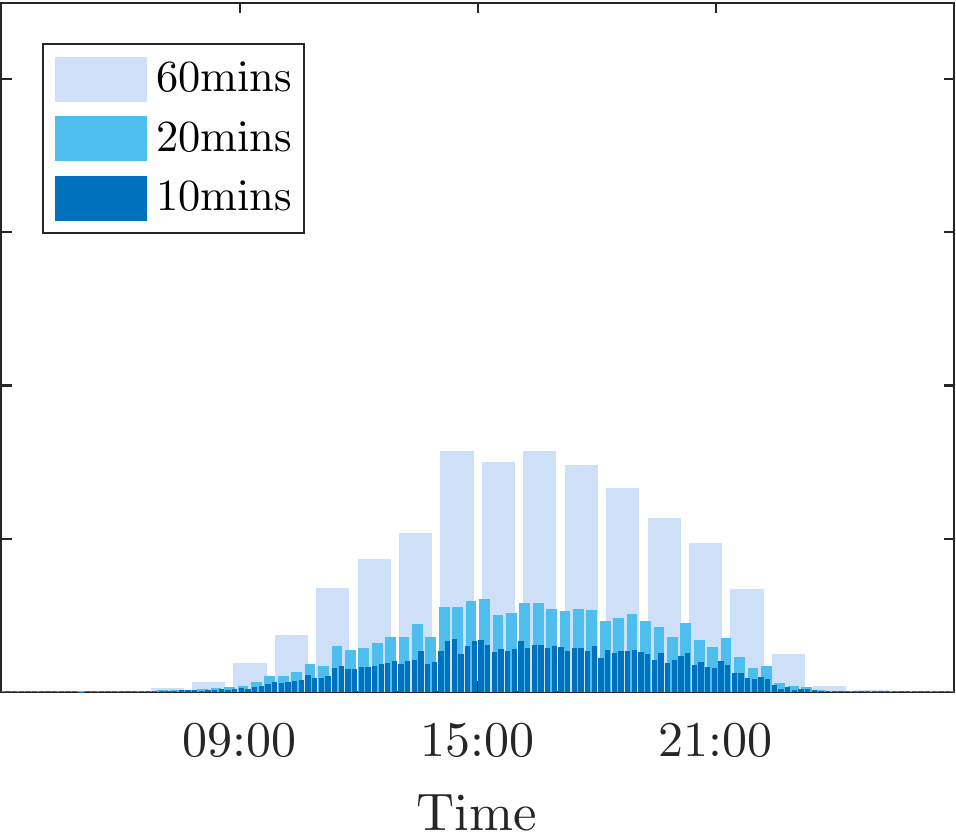} \\
	\end{tabular}
	\caption{\textbf{Number of pedestrians (unique devices) detected at different hours of the day} at the entrances to three buildings ($A$, $F$, and $G$) in the overpass system from 21 (Monday) to 27 (Sunday) November 2016.}
	\label{fig_WeeklyLocationDensity}
\end{figure*}

\rev{
\textbf{Identifying transportation modes:} Four transportation modes (non-vehicle, metro, bus, and car) are identified using location values and timestamps. Non-vehicle mode is interpolated as walking for cycling is not a common commuting tool in Singapore. Data points belonging to a PoI are identified as non-vehicle mode. For data points of a trip, the identification is done in a hierarchical module as shown in Fig.~\ref{fig_data_collection_preproseccing}. Two point-wise features used for vehicle-or-not classifier are moving-averaged speed and acceleration. Totally 16 segment-wise features are used in the vehicle type classifier. The most important three are:
\begin{enumerate*}
  \item the 85th percentile of speed of data points in this segment,
  \item mean value of speed of data points in this segment,
  \item time that the person waited before this vehicle segment started.
\end{enumerate*}
After trials and errors, the adopted vehicle-or-not classifier is an Adaptive Boosting classifier and the adopted vehicle type classifier is a Random Forest classifier. The final point-wise mode identification accuracy is 81 percent.}

After finishing the above steps, investigation can be conducted to get deeper insights. Example results are shown in Fig.~\ref{fig_NSE_results} which is generated using 2015 data. In Fig.~\ref{fig_home_school}, the heavily crossed links between homes and schools indicate that many students travel longer distance to schools which are not the closest ones to their homes. This can be interpreted as that geographic closeness is not the dominant factor for school choosing. From Fig.~\ref{fig_mode_frac}, we can find that:
\begin {enumerate*}
\item When home and school are close, most students prefer to walk rather than to ride vehicles;
\item For distance which is not suitable for walking but still not too long, bus is more popular;
\item When the home-school distance keeps increasing, students ride more metros than buses;
\item Car usage stays relatively stable for different levels of home-school distance.
\end {enumerate*}


\section{Building-Wide Case Study}

\revtwo{In this case study, passive WiFi tracking (by collecting WFPRRs) is adopted to investigate pedestrian visiting and moving behaviors between buildings without influencing them. The studied area is a busy public district consisting of shopping malls, office buildings, and a transportation hub. Precise insights of pedestrian movement are obtained.}

\subsection{System Setup and Data Pipeline}

The core of the data collection system is gateway node (GN) which sniffs the probe requests sent out by passing-by pedestrians' mobile devices. 
In this project, Raspberry Pi Micro Processor is used as the GN.
The studied district consists of seven buildings ($A$-$G$) connected to each other by sheltered overpass on the second floor, as shown in Fig.~\ref{fig_mobility}.
At the entrance to each building in this overpass system, multiple GNs are deployed to ensure the full coverage of receiving probe requests and detecting passing-by pedestrians. 

The entire data pipeline consists of collection, uploading, preprocessing, and analysis.
In data collection, GN adds additional information with received probe request, including check-in time, check-out time, signal strength, and MAC address. Data are uploaded and stored in the cloud. In data preprocessing, we combine GNs in the same building using "OR" relationship and remove repeated MAC addresses in the same building within the same time frame. Afterwards, by connecting records with the same MAC address, we are able to catch pedestrian paths and thus further insights.

\subsection{Analysis and Outcomes}

\begin{figure}[!b]
\centering
\includegraphics[width=\linewidth]{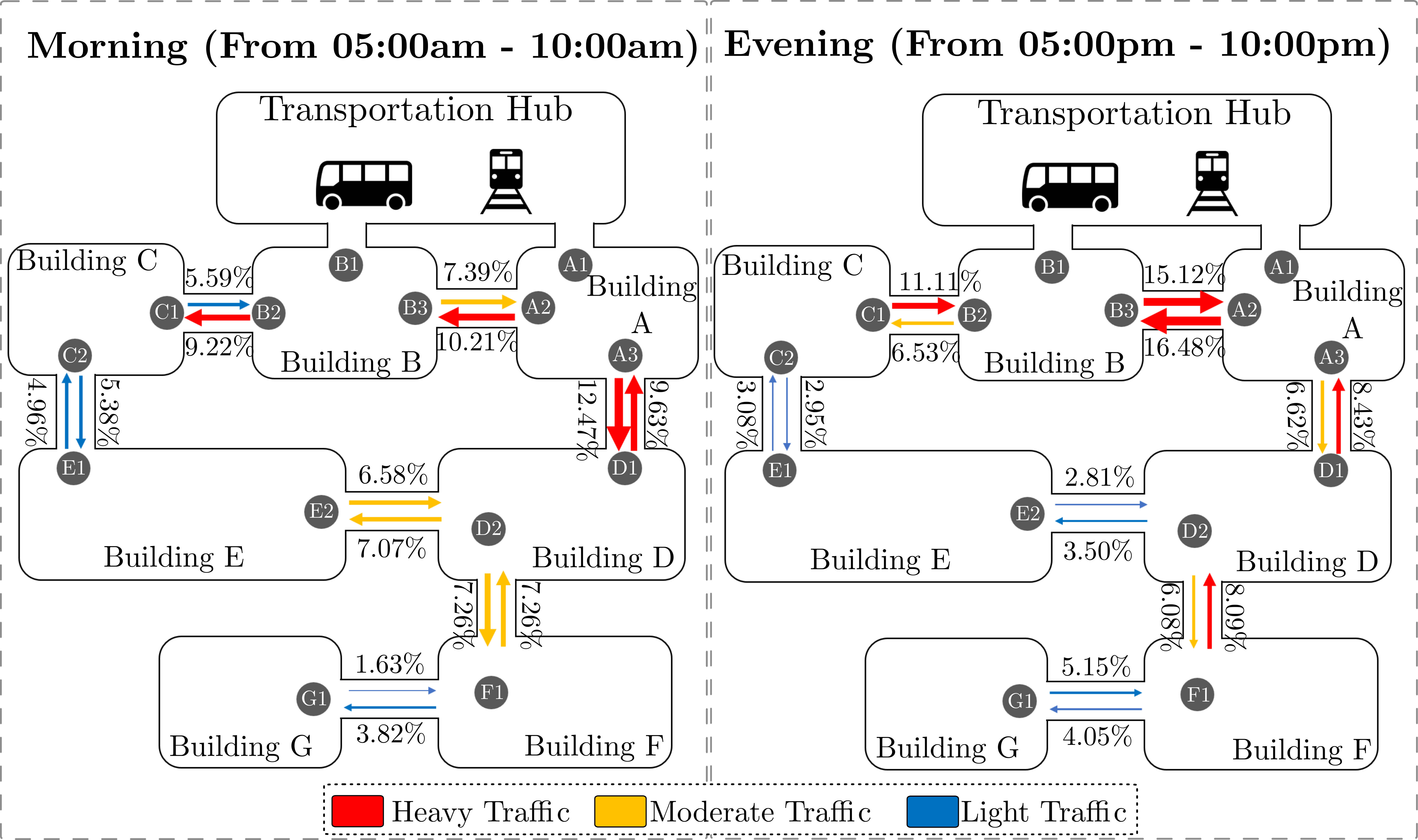}
\caption{\textbf{Pedestrian flow density between buildings} of the studied district in two different time periods on 5 January 2017 (Thursday). Links between buildings represent overpasses. Percentage value besides each arrow represents the traffic occupied by that particular direction among the overpass system.}
\label{fig_mobility}
\end{figure}

We first look into people's visiting statistics to particular buildings. Figure~\ref{fig_WeeklyLocationDensity} visualizes one-week device detection data of building $A$, $F$, and $G$. On weekdays, building $A$ had the daily peak at night and the weekly peak on Friday night. This can be explained by that building $A$ is a shopping mall and is located near to the transportation hub, as commutators tend to go dining or shopping after work. In contrast, building $F$ is a functional building with people working in it, so it had three daily peaks corresponding to arriving to work, lunch break, and leaving after work. Building $G$ has fewer devices detected compared with $A$ and $F$. Because it is located furthest from the transportation hub and at the end of the overpass system, fewer people reach it through the overpass system.

\rev{Next, we investigate pedestrian flows in the overpass system as presented in Fig.~\ref{fig_mobility}. From the figure, more people traveled downwards in the morning (i.e. from $A$ and $B$ to other buildings), while the reverse happened in the evening. This can be understood as crowds coming from transportation hub to the buildings for different purposes (e.g. working). Moreover, compared with morning, movement between Building $C, D$, and $E$ was less active in the evening, while movement between Building $A, B$, and $C$ was more active in the evening. This can be explained by the building functions. Building $D$ and $E$ are normal functional buildings, so more people are around during working hour. Building $A, B$, and $C$ are shopping malls, so more people are around after working hour.}



\begin{figure}[!t]
\centering
\begin{tabular}{cc}
(a) C1 in, C1 out  & (b) C2 in, C1 out \\ 
\includegraphics[width=0.22\textwidth]{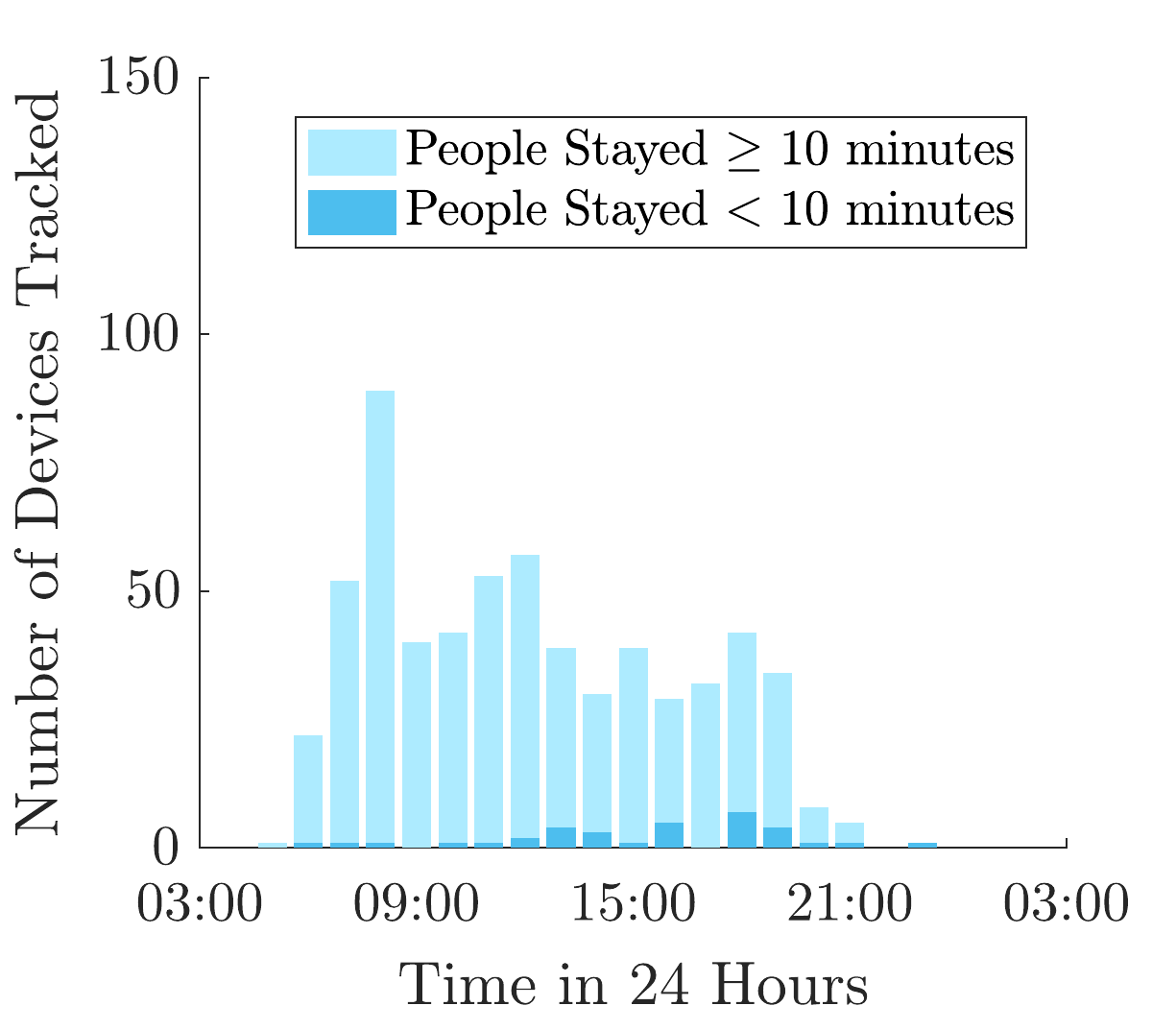}	& 
\includegraphics[width=0.22\textwidth]{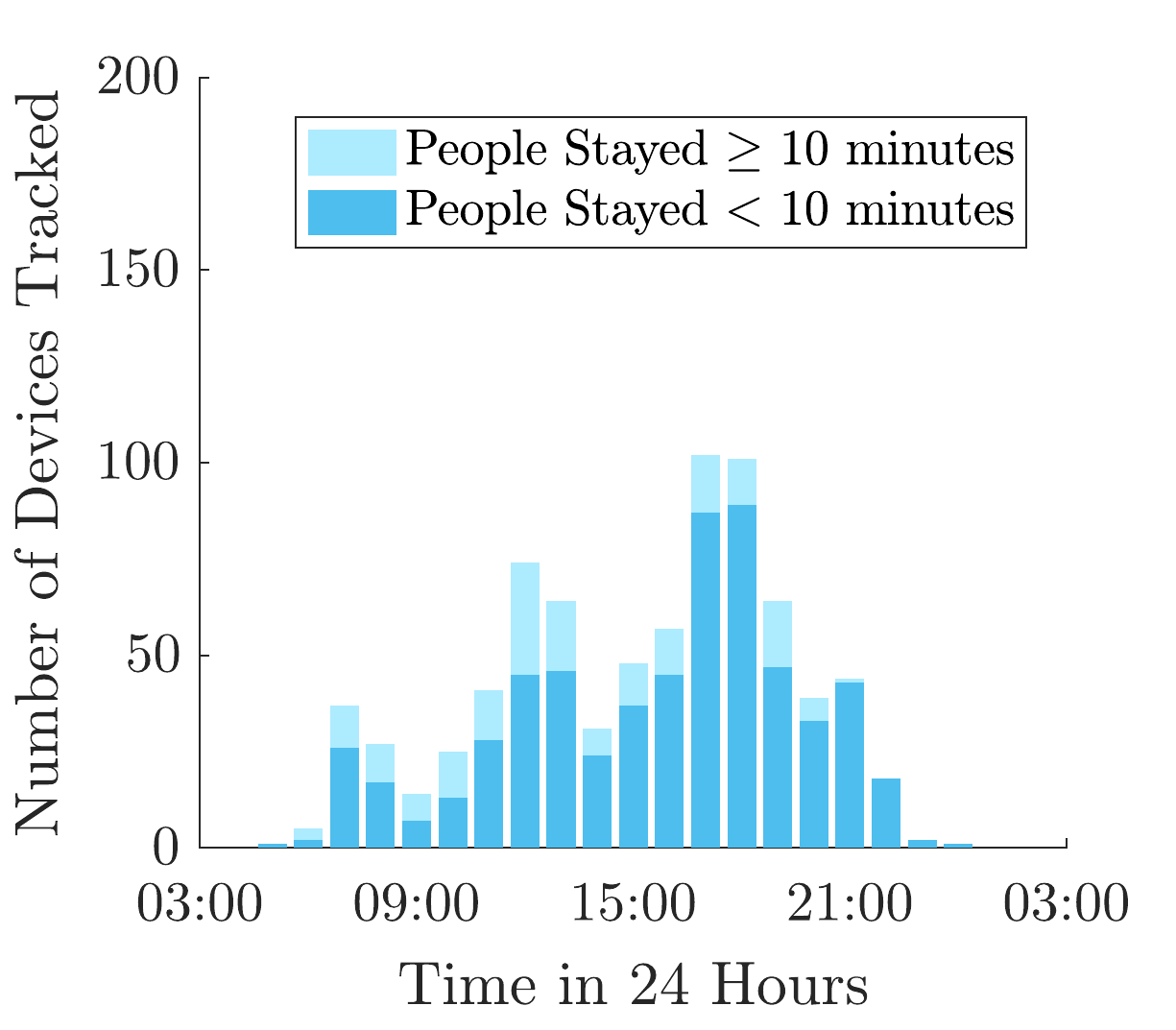} \\
(c) C1 in, C2 out & (d) C2 in, C2 out\\ 
\includegraphics[width=0.22\textwidth]{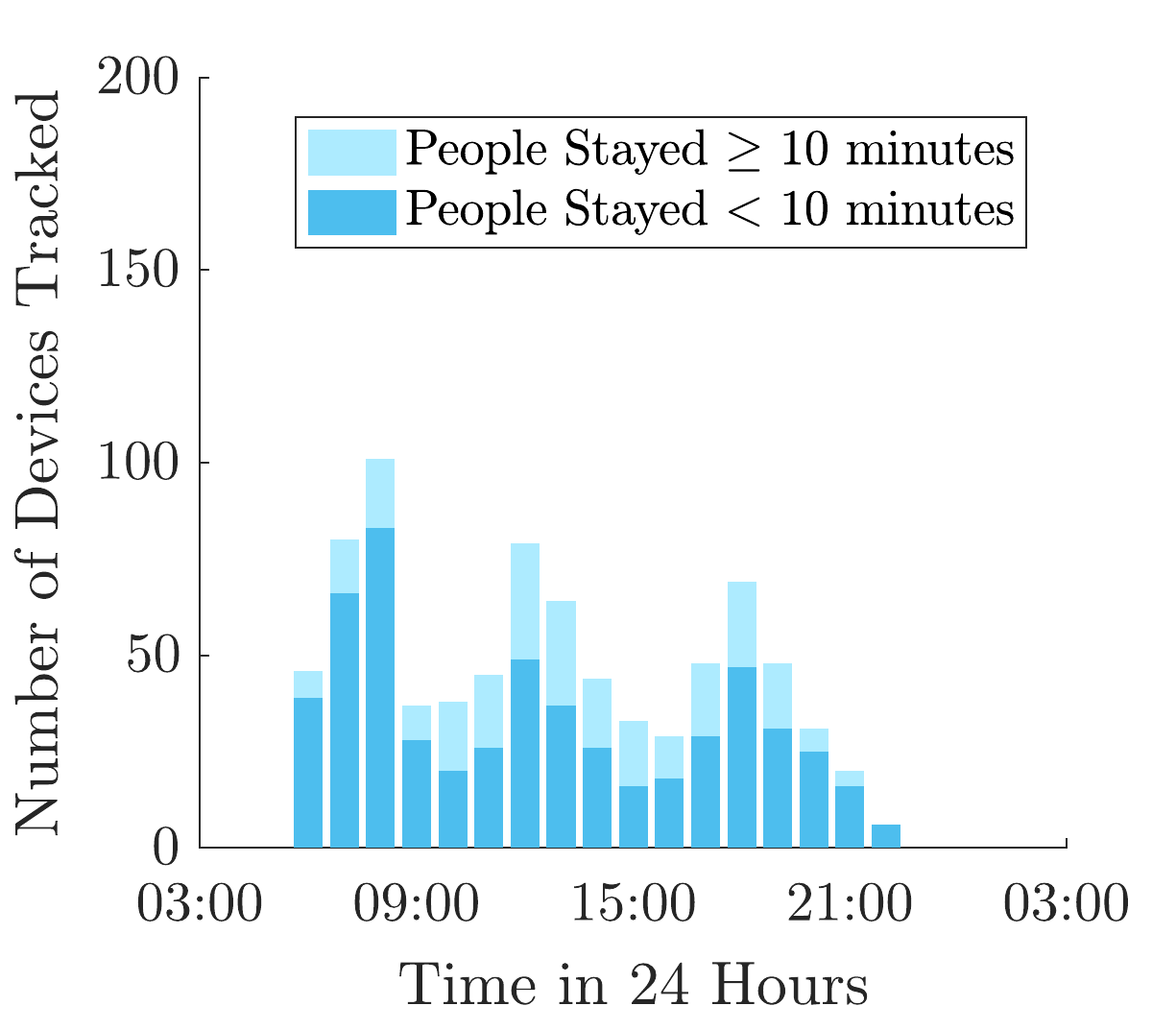}	& 
\includegraphics[width=0.22\textwidth]{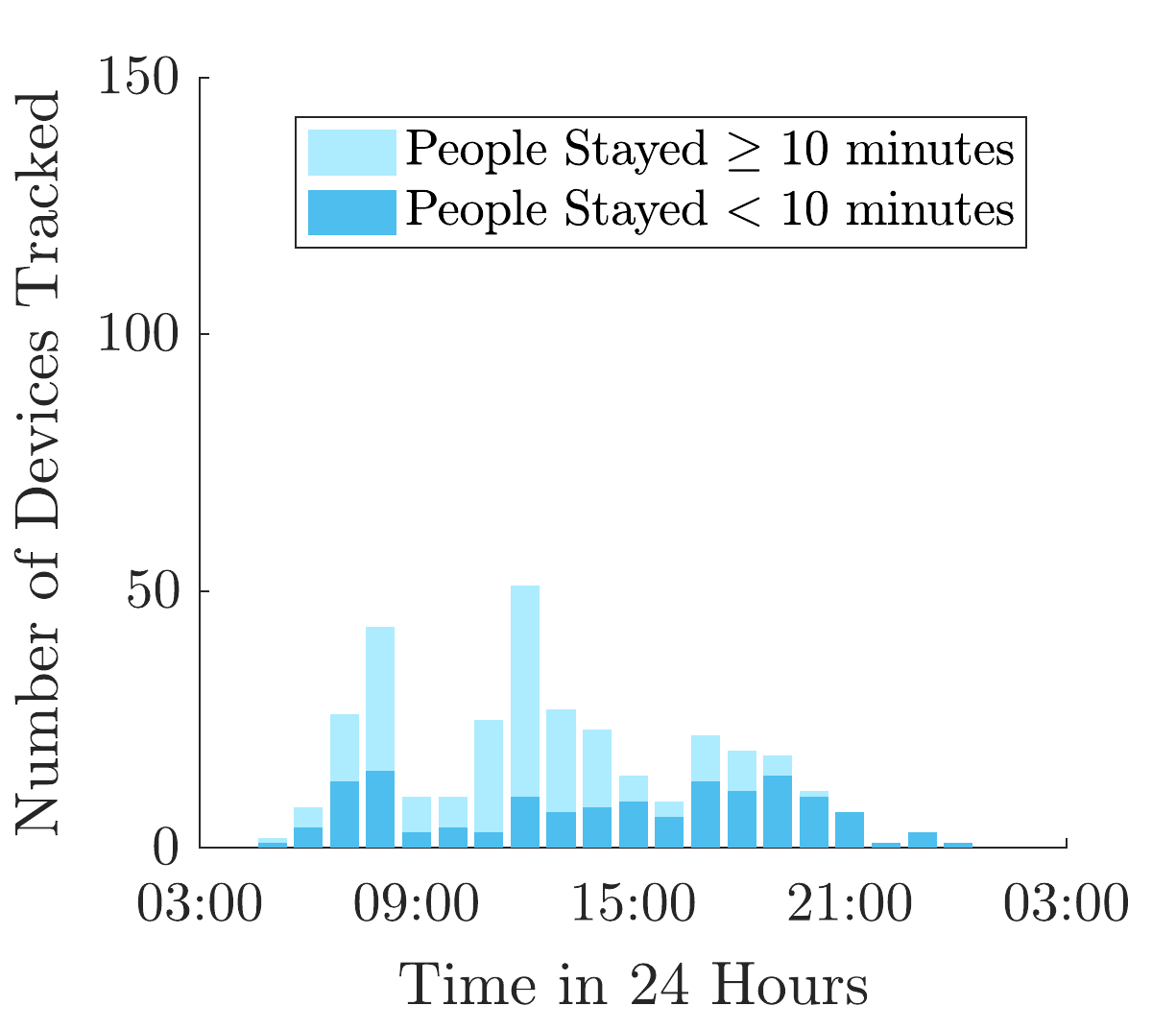}
\end{tabular}
\caption{\textbf{Counts of pedestrians entering and exiting building $C$} through two entrances ($C$1 and $C$2) on 28 November 2016 (Monday). Each bar is broken down into two different staying durations as shown in the legends.
}
\label{fig_stayTime}
\end{figure}

Finally, we scale down to study pedestrians entering and existing one single building. Shopping mall building $C$ is taken as an example which has two entrances as shown in Fig.~\ref{fig_mobility}.
The related statistics are visualized in Fig.~\ref{fig_stayTime}. 
From the figure, people who entered $C$ through entrance $C$1 and exited through $C$2 or the reverse mostly did not stay for more than 10 minutes. This means that pedestrians used building $C$ as a passing-through route. On the contrary, most of the pedestrians who entered and exited through the same entrance $C$1 stayed in $C$ for a long time. Counts of people who entered and exited $C$ through the same entrance $C$2 have three peaks at 09:00am, 01:00pm, and 06:00pm which can be explained by people coming from other buildings to have meals in $C$. Moreover, most people who entered building $C$ through $C$2 in the afternoon had passing-through purpose.

\section{Conclusion}

In this paper, we first summarize recent studies of understanding urban human mobility through crowdsensed data by seven common data types and six popular human mobility subjects. Afterwards, a tutorial is presented to discuss pros and cons of each data type, details of each human mobility subject, fitness of each data type to each mobility subject, and common data preprocessing and analysis methods.
Finally, two case studies are presented. The city-wide case study leverages WiFi-based localization to study mobility patterns of students in a very large scale. Fine-grained mobility patterns are obtained from only SWFAPRs data. The building-wide project collects mobile devices' WiFi probe requests passively to gain insights of building-wide pedestrian mobility in a representative public district.

\section*{Acknowledgments}
This work was supported in part by the National Research Foundation, Prime Minister’s Office, Singapore under its National Science Experiment project (Grant RGNRF1402), and in part by the Singapore Ministry of National Development (MND) Sustainable Urban Living Program under Grant SUL2013-5. We would also like to thank all the involved researchers, interns, and collaborators.


\bibliographystyle{IEEEtran}
\bibliography{main}

\end{document}